\documentclass{emulateapj}

\usepackage{verbatim}
\usepackage{natbib}
\usepackage{amsmath}
\usepackage{amsbsy}
\usepackage{lscape}
\usepackage{epstopdf}

\slugcomment{Submitted to ApJ and revised following the referee's report}
\shorttitle{Implications of Intergalactic Opacity}
\shortauthors{Faucher-Gigu\`ere et al.}

\newcommand{\Lya}{\mbox{Ly$\alpha$}}

\begin{document}

\title{Evolution of the Intergalactic Opacity: Implications for the Ionizing Background, Cosmic Star Formation, and Quasar Activity}

\author{Claude-Andr\'e Faucher-Gigu\`ere\altaffilmark{1}, Adam Lidz\altaffilmark{1}, Lars Hernquist\altaffilmark{1}, Matias Zaldarriaga\altaffilmark{1,2}}
\altaffiltext{1}{Department of Astronomy, Harvard University, Cambridge, MA, 02138, USA; cgiguere@cfa.harvard.edu.}
\altaffiltext{2}{Jefferson Physical Laboratory, Harvard University, Cambridge, MA, 02138, USA.}

\begin{abstract}
We investigate the implications of the intergalactic opacity for the evolution of the cosmic ultraviolet luminosity density and its sources.
Our main constraint follows from our measurement of the Lyman-$\alpha$ forest opacity at redshifts $2\leq z\leq4.2$ from a sample of 86 high-resolution quasar spectra.
In addition, we impose the requirements that intergalactic HI must be reionized by $z=6$ and HeII by $z\approx3$, and consider estimates of the hardness of the ionizing background from HI to HeII column density ratio measurements.
The derived hydrogen photoionization rate 
is remarkably flat over the \Lya~forest redshift range covered.
Because the quasar luminosity function is strongly peaked near $z\sim2$, the lack of redshift evolution indicates that star-forming galaxies likely dominate the photoionization rate at $z\gtrsim3$, and possibly at all redshifts probed.
Combined with direct measurements of the galaxy UV luminosity function, this requires only a small (emissivity weighted) fraction $f_{\rm esc}\sim0.5\%$ of galactic hydrogen ionizing photons to escape their source for galaxies to solely account for the entire ionizing background.
Under the assumption that the galactic UV emissivity traces the star formation rate (which is the case if the escape fraction and dust obscuration are constant with redshift), current state-of-the-art observational estimates of the star formation rate density, which peak similarly to quasars at $z\sim2$, appear to underestimate the total photoionization rate at $z\approx4$ by a factor $\approx4$, are in tension with the most recent determinations of the UV luminosity function, and fail to reionize the Universe by $z=6$ if extrapolated to arbitrarily high redshift.
A star formation history peaking earlier, as in the theoretical model of Hernquist \& Springel, fits the \Lya~forest photoionization rate well, reionizes the Universe in time, and is in better agreement with the rate of $z\approx4$ gamma-ray bursts observed by \emph{Swift}.
Quasars suffice to doubly ionize helium by $z\approx3$, provided that most of the HeII ionizing photons they produce escape into the intergalactic medium, and likely contribute a non-negligible and perhaps dominant fraction of the hydrogen ionizing background at their $z\sim2$ peak.
\end{abstract}


\section{INTRODUCTION}
\label{introduction}
The opacity of the Lyman-$\alpha$ (\Lya) forest is set by a competition between hydrogen photoionizations and recombinations \citep[][]{1965ApJ...142.1633G} and can thus serve as a probe of the photonization rate \cite[e.g.][]{1997ApJ...489....7R}.
The hydrogen photoionization rate $\Gamma$~is a particularly valuable quantity as it is an integral over all sources of ultraviolet (UV) radiation in the Universe,
\begin{equation}
\Gamma(z) = 
4\pi \int_{\nu_{\rm HI}}^{\infty} 
\frac{d\nu}{h \nu}
 J_{\nu}(z) \sigma(\nu),
\end{equation}
where $J_{\nu}$ is the angle-averaged specific intensity of the background, $\sigma(\nu)$ is the photoionization cross section of hydrogen, and the integral is from the Lyman limit to infinity.
As such, it bears a signature of cosmic stellar and quasistellar activity that is not subject to the completeness issues to which direct source counts \citep[e.g.,][]{1999ApJ...519....1S, 2004ApJ...606L..25B, 2006ApJ...642..653S, 2007ApJ...670..928B, 2006ApJ...653..988Y, 2007ApJ...654..731H} are prone.
In addition, unlike the radiation backgrounds directly observed on Earth, the \Lya~forest is a \emph{local} probe of the high-redshift UV background, as only sources at approximately the same redshift contribute to $\Gamma$~at any point in the forest \citep[][]{1996ApJ...461...20H}.

The hydrogen photoionization rate is not only a powerful observational tracer of cosmic radiative history, but is also a key ingredient in cosmological simulations and theoretical models of galaxy formation.
Photoionization heating can, for example, suppress the formation of dwarf galaxies \citep[][]{1992MNRAS.256P..43E, 1996ApJ...465..608T, 1997ApJ...477....8W}.
While the photoionization rate of hydrogen by itself provides limited information on the shape of the background spectrum, it can be combined with the photoionization rates of other species (helium, for example) to provide a direct measure of the background hardness \citep[e.g.,][]{1997ApJ...488..532C, 2003MNRAS.340..473S, 2006MNRAS.366.1378B}.
The background spectrum sets the temperature of the gas and so it must be specified in any hydrodynamical simulation of structure formation \citep[e.g.,][]{1996ApJ...457L..51H, 1999ApJ...511..521D, 2003MNRAS.339..312S}.
Knowledge of the ionizing background spectrum is also necessary to infer the mass density of metals from observations of absorption by particular ions (``ionization corrections'') \citep[e.g.,][]{2003ApJ...596..768S, 2004ApJ...602...38A, 2007arXiv0712.1239A}.

Recent results make it especially worthwhile to reconsider the implications for the ionizing sources, usually assumed to be a combination of quasars and star-forming galaxies \citep[e.g.,][]{1999ApJ...514..648M}, of the \Lya~forest opacity in the light of new data.
The comoving star formation rate (SFR) density as estimated from the observed luminosity of galaxies, for instance, is in tension with the independently derived evolution of the stellar mass density \citep[e.g.,][]{2006ApJ...651..142H}.
This suggests problems with observational estimates of the cosmic SFR, to which $\Gamma$~is sensitive.
The abundance of intergalactic metals as traced by CIV appears consistent with being constant from $z\sim2$ to perhaps $z\sim6$ \citep[][]{2001ApJ...561L.153S, 2003ApJ...594..695P, 2006MNRAS.371L..78R}, suggestive of an early period of nucleosynthetic activity that may be at odds with estimates of the SFR density that decline above $z\sim2-3$ \citep[][]{2004ApJ...615..209H, 2006ApJ...651..142H}.
Stellar mass density measurements at $z\gtrsim5$ \citep[e.g.,][]{2006ApJ...649L..67L, 2006ApJ...651...24Y, 2007MNRAS.374..910E, 2007ApJ...659...84S} also suggest significant early star formation. 
In addition, deep $z\sim6-10$ searches utilizing magnification by foreground galaxy clusters \citep[][]{2006A&A...456..861R, 2007ApJ...663...10S} have yielded surprisingly large abundances of low-luminosity galaxies, in tension with blank field surveys and perhaps even requiring an upturn in their abundance with increasing redshift.
At lower redshifts, \cite{2007arXiv0711.1354R} discovered a large population of faint \Lya~emitters at $2.67\leq z \leq 3.75$ in an extremely deep 92-hour exposure with VLT/FORS2, most of which would not have been selected in existing Lyman-break surveys.
This underscores the definite possibility that photometric-selection surveys may presently miss a substantial fraction of the cosmic luminosity density originating in low-luminosity objects.

Here, we use improved measurements of the intergalactic opacity at $z\gtrsim2$ to obtain a measurement of the hydrogen photoionization rate $\Gamma$.
We present a detailed comparison with direct source counts of quasars and star-forming galaxies and ask whether these can account for the measured $\Gamma$.
Finally, we consider the implications for the estimates of the cosmic SFR density and the escape fraction of ionizing photons from galaxies, which relates the soft UV emission to the photons that actually contribute to photoionizations.

The observational constraints we use are described in \S \ref{measurements}.
In \S \ref{gamma}, we derive the hydrogen photoionization rate implied by our recent measurement of the \Lya~forest opacity \citep[][]{2008ApJ...681..831F} and explore potential systematic effects.
In \S \ref{photoionizations to emissivity}, we describe how to convert the photoionization rate to a UV luminosity density.
We investigate the implications for the cosmic sources of UV radiation, quasars and star-forming galaxies in \S \ref{cosmic sources}.
We also consider the constraints imposed by the reionizations of hydrogen and helium in this section.
We revisit the star formation history in \S \ref{madau diagram} before discussing our results in \S \ref{discussion} and summarizing our conclusions \S \ref{conclusions}.

Throughout, we assume a cosmology with $(\Omega_{\rm m},~\Omega_{\rm b},~\Omega_{\Lambda},~h,~\sigma_{8})=(0.28,~0.046,~0.72,~0.70,~0.82)$, as inferred from the \emph{Wilkinson Microwave Anisotropy Probe} (WMAP) five-year data in combination with baryon acoustic oscillations and supernovae \citep[][]{2008arXiv0803.0547K}.
Unless otherwise stated, all error bars are $1\sigma$.
Some of the results presented here were already reported in a concise companion \emph{Letter} \citep[][]{2008ApJ...682L...9F}.
They are described here with the full details of the analysis and underlying assumptions, as well as supporting arguments.

\begin{deluxetable}{lcccccl}
\tablewidth{0pc}
\tablecaption{\Lya~Effective Optical Depth and Derived Hydrogen Photoionization Rate\label{our measurement}}
\tabletypesize{\footnotesize}
\tablehead{\colhead{$z$} & \colhead{$\tau_{\rm eff}$}\tablenotemark{a} & \colhead{$\sigma_{\tau_{\rm eff}}$}\tablenotemark{b} & $\Gamma$\tablenotemark{c} & $\sigma_{\Gamma}$\tablenotemark{d} \\
\colhead{} & \colhead{} & \colhead{} & 10$^{-12}$ s$^{-1}$ & 10$^{-12}$ s$^{-1}$}
\startdata
2.0 & 0.127 & 0.018 & 0.64 & 0.18 \\
2.2 & 0.164 & 0.013 & 0.51 & 0.10 \\
2.4 & 0.203 & 0.009 & 0.50 & 0.08 \\
2.6 & 0.251 & 0.010 & 0.51 & 0.07 \\
2.8 & 0.325 & 0.012 & 0.51 & 0.06 \\
3.0 & 0.386 & 0.014 & 0.59 & 0.07 \\
3.2 & 0.415 & 0.017 & 0.66 & 0.08 \\
3.4 & 0.570 & 0.019 & 0.53 & 0.05 \\
3.6 & 0.716 & 0.030 & 0.49 & 0.05 \\
3.8 & 0.832 & 0.025 & 0.51 & 0.04 \\
4.0 & 0.934 & 0.037 & 0.55 & 0.05 \\
4.2 & 1.061 & 0.091 & 0.52 & 0.08 \\
\enddata
\tablenotetext{a}{Measurement reported by \cite{2008ApJ...681..831F} corrected for continuum bias binned in $\Delta z=0.2$ intervals and with the contribution of metals to absorption in the \Lya~forest subtracted based on the results of \cite{2003ApJ...596..768S}.}
\tablenotetext{b}{Statistical only.}
\tablenotetext{c}{Derived as in \S \ref{gamma}.}
\tablenotetext{d}{Includes contributions from the statistical uncertainties in both $\tau_{\rm eff}$ and $T_{0}$.}
\end{deluxetable} 

\section{OBSERVATIONAL CONSTRAINTS}
\label{measurements}
The main observational constraint we use is our measurement of the \Lya~forest effective optical depth between $z=2$ and $z=4.2$ \citep[][]{2008ApJ...681..831F}.
This measurement, based on 86 quasar spectra obtained with the HIRES and ESI spectrographs on Keck and with MIKE on Magellan, is the most precise to date from high-resolution, high-signal-to-noise data in this redshift interval.
We use our measurement corrected for continuum bias binned in $\Delta z=0.2$ intervals with the contribution of metals to absorption in the forest subtracted following the results of \cite{2003ApJ...596..768S}.
For convenience, these data points are reproduced in Table \ref{our measurement}.
In \cite{2008ApJ...681..831F}, we provided estimates of the systematic uncertainty in $\tau_{\rm eff}$ arising from the continuum and metal corrections.
In this paper, we are however most concerned with the derived photoionization rate.
As the systematic uncertainties in converting $\tau_{\rm eff}$ to $\Gamma$ (involving the thermal history of the IGM) are larger than those from the continuum and metal corrections, we simply use the statistical error bars on $\tau_{\rm eff}$ and separately explore the systematic effects influencing our inference of $\Gamma$ in this work.

We further employ the constraints set by the reionizations of hydrogen and helium.
Specifically, we use the fact that hydrogen is reionized by $z=6$, as indicated by the absence of a complete \cite{1965ApJ...142.1633G} trough in the spectra of quasars at this redshift \citep[see, e.g.][for a review of the present observational constraints on reionization]{2006ARA&A..44..415F}.
We assume that helium is doubly ionized by $z\approx3$, as supported by UV spectra of the HeII \Lya~forest of $z\sim3$ quasars.
\citep[]{1994Natur.370...35J, 1997A&A...327..890R, 1997AJ....113.1495H, 2000ApJ...534...69H, 2002ApJ...564..542S, 2005A&A...442...63R, 2006A&A...455...91F}.

Finally, we use estimates of the spectrum of the softness of the UV background \citep[][]{2006MNRAS.366.1378B} from measurements of the HI to HeII column density ratio obtained by combining HI and HeII \Lya~forest spectra \citep[][]{2004ApJ...605..631Z, 2004ApJ...600..570S}. 

\section{THE PHOTOIONIZATION RATE FROM THE \Lya~FOREST}
\label{gamma}
\begin{figure}[ht]
\begin{center}
\includegraphics[width=0.49\textwidth]{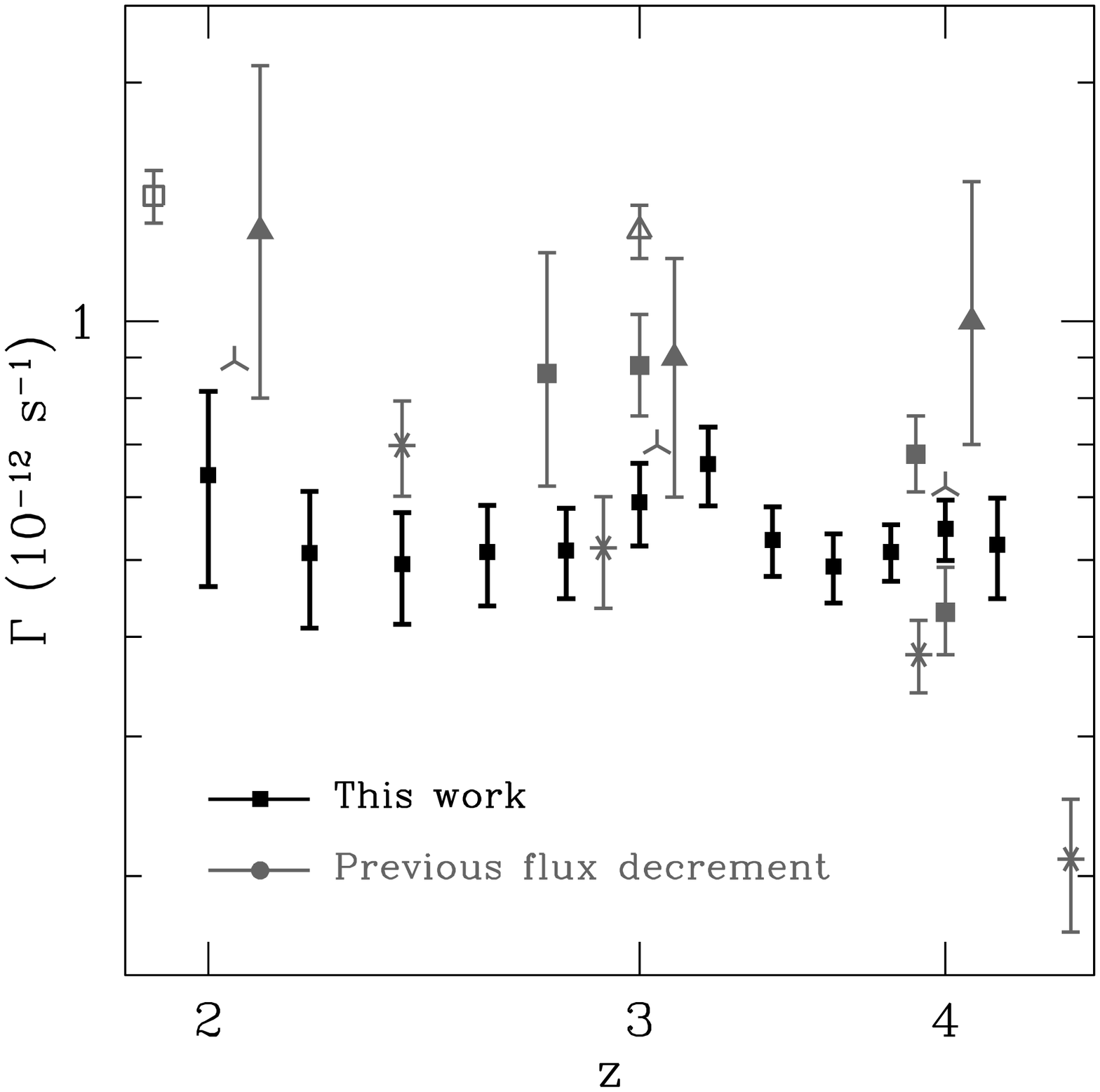}
\end{center}
\caption{Intergalactic hydrogen photoionization rate inferred from our \Lya~effective optical depth measurement (black squares) compared with existing measurements using a similar flux decrement method.
The previous measurements are shown in gray: \cite{1997ApJ...489....7R} (three-legged asterisks), \cite{2001ApJ...549L..11M} (asterisks), \cite{2004MNRAS.350.1107M} (solid squares), \cite{2004ApJ...617....1T} (open square), \cite{2005MNRAS.357.1178B} (solid triangles), and \cite{2005MNRAS.360.1373K} (open triangle).
Some of the previous measurements have been slightly offset in redshift in order to improve readability of the plot.}
\label{gamma comparison}
\end{figure}

We begin by deriving the hydrogen photoionization rate $\Gamma$ implied our measurement of the \Lya~forest opacity.
The method we employ is based on the mean level of absorption in the forest and is termed the ``flux decrement'' method \citep[][]{1997ApJ...489....7R}.
For the reader who is not concerned with the details, we immediately present the results in Figure \ref{gamma comparison}, with numerical values tabulated in Table \ref{our measurement}.
The figure compares our measurement to previous estimates based on the flux decrement method that overlap in redshift \citep{1997ApJ...489....7R, 2001ApJ...549L..11M, 2004MNRAS.350.1107M, 2004ApJ...617....1T,2005MNRAS.357.1178B, 2005MNRAS.360.1373K}. 
We refer to the appendix of \cite{2008ApJ...673...39F} for a detailed numerical summary of the previous $\Gamma$ measurements and to \S \ref{previous work} for a discussion.

\subsection{\Lya~Effective Optical Depth}
\label{lya effective optical depth}
The \Lya~effective optical depth is defined as
\begin{equation}
\label{taueff definition}
\tau_{\rm eff}(z) \equiv -\ln{[\langle F \rangle(z)]},
\end{equation}
where $\langle F \rangle(z)$ is the mean transmission of the \Lya~forest at redshift $z$.
Neglecting redshift-space distortions,
\begin{equation}
F = \exp{(-\tau)},
\end{equation}
where
\begin{equation}
\label{tau gp}
\tau = \frac{\pi e^{2} f_{\Lya}}{m_{e} \nu_{\Lya}}
\frac{1}{H(z)}
\frac{R(T)n_{\rm HII} n_{e}}{\Gamma}
\end{equation}
is the local \cite{1965ApJ...142.1633G} optical depth.
Here, $e$ and $m_{e}$ are the charge and mass of the electron, $f_{\Lya}$ and $\nu_{\Lya}$ are the \Lya~oscillator strength and frequency, $H(z)$ is the Hubble parameter, $R(T)$ is the temperature-dependent hydrogen recombination rate, $n_{\rm HII}$ and $n_{e}$ are the proper number densities of HII and electrons, and $\Gamma$ is the hydrogen photoionization rate.

We consider a gas made purely of hydrogen and helium, with mass fractions $X$ and $Y$, respectively.
To describe its ionization state, we define $(x, y_{\rm II}, y_{\rm III}) \equiv (n_{\rm HII}/n_{\rm H}, n_{\rm HeII}/n_{\rm He}, n_{\rm HeIII}/n_{\rm He})$.
Then
\begin{equation}
n_{\rm HII} = \frac{\rho_{crit}\Omega_{b}}{m_{p}} X x (1+\delta) (1+z)^{3}
\end{equation}
and
\begin{equation}
\label{n electrons}
n_{e} = \frac{\rho_{crit}\Omega_{b}}{m_{p}}[ X x + 0.25Y(y_{\rm II} + 2 y_{\rm III})](1+\delta) (1+z)^{3},
\end{equation}
where $\delta$ is the local overdensity, and we have used the fact that a helium atom weighs approximately $4m_{p}$.
To a good approximation,
\begin{equation}
\label{h recombination coefficient}
R(T) \approx R_{0} T^{-0.7},
\end{equation}
with $R_{0}=4.2 \times 10^{-13} {\rm~cm}^{3}~{\rm s}^{-1}/(10^{4} {\rm~K})^{-0.7}$
\citep[e.g.,][]{1997MNRAS.292...27H}, and the Hubble parameter is given by the Friedmann equation,
\begin{equation}
\label{friedmann eq}
H(z)=H_{0} \sqrt{\Omega_{m}(1+z)^{3} + \Omega_{\Lambda}}.
\end{equation}
Assuming that the intergalactic gas follows a power-law temperature-density relation of the form
\begin{equation}
\label{temperature density relation}
T=T_{0}(1+z)^{\beta}
\end{equation}
\citep[][]{1997MNRAS.292...27H}, equation (\ref{tau gp}) implies that
\begin{equation}
\tau = A(z)(1+\delta)^{2 - 0.7\beta},
\end{equation}
with
%\begin{equation}
\begin{align}
\label{a definition}
A(z) & \equiv 
\frac{\pi e^{2} f_{\Lya}}{m_{e} \nu_{\Lya}}
\left( \frac{\rho_{crit} \Omega_{b}}{m_{p}} \right)^{2}
\frac{1}{H(z)} \notag \\
& X x [X x + 0.25Y(y_{\rm II} + 2 y_{\rm III})] 
\frac{R_{0} T_{0}^{-0.7}}{\Gamma}
(1 + z)^{6}.
\end{align}
%\end{equation}
Note that equation (\ref{temperature density relation}) is often expressed in term of the parameter $\gamma=\beta+1$ in the literature.
An isothermal temperature-density relation corresponds to $\gamma=1$ or, equivalently, $\beta=0$.

Given a probability density function (PDF) for the gas density $\Delta \equiv 1 + \delta$,
\begin{equation}
\langle F \rangle(z) =
\int_{0}^{\infty}
d\Delta
P(\Delta; z)
\exp{(-\tau)}.
\end{equation}
Note that neglecting redshift-space distortions from thermal broadening and peculiar velocities is well-motivated here since $\langle F \rangle$ is to first order independent of the smoothing of the $F$ field induced by these distortions.\footnote{The mean of $F$ is not strictly independent of redshift space distortions, since these really smooth the $\tau$ field, of which $F$ is a nonlinear function.
This however amounts to a small effect.}
For any specified $\beta$, a measurement of $\tau_{\rm eff}(z)$ thus determines the evolution of $A(z)$, which itself fixes the ratio $T_{0}^{-0.7}/\Gamma$ if the cosmological parameters and ionization state of the gas are known (eq. \ref{a definition}).

\subsection{Gas Density PDF}
\label{gas density pdf}
\cite{2000ApJ...530....1M} derived an approximate analytical functional form for the volume-weighted gas density PDF,
\begin{equation}
P_{V}(\Delta) = 
A \exp{\left[ 
-\frac{(\Delta^{-2/3} - C_{0})^{2}}{2(2\delta_{0}/3)^{2}} \Delta^{-b}
\right]}.
\end{equation}
This formula is motivated by the facts that the gas in the low-density voids, where gravity is negligible, should expand at nearly constant velocity, whereas the distribution in high-density collapsed structures should be determined by their power-law profile, $\Delta \propto r^{-3/(b-1)}$.\footnote{Note that \cite{2000ApJ...530....1M} use the notation $\beta$ for $b$.
We have altered their notation in order to avoid confusion with the temperature-relation slope parameter defined in this work (eq. \ref{temperature density relation}).}

The parameters $A$ and $C_{0}$ are fixed by requiring the total volume and mass probabilities to be normalized to unity.
Fits to a numerical simulation at $z=$2, 3, and 4 yields the parameters given in Table \ref{MHR params}, with $\delta_{0}=7.61/(1+z)$; at $z=6$, $b=2.5$, corresponding to isothermal halos, is assumed.

\cite{1997ApJ...489....7R} compared the density distributions from simulations using very different numerical methods (Eulerian versus Lagrangian smoothed particle hydrodynamics) and cosmologies ($\Lambda$CDM vs. Einstein-de Sitter).
They found excellent agreement in spite of the different simulation parameters.
\cite{2000ApJ...530....1M} attributed this similarity to the fact that the density distribution should depend mostly on the amplitude of the density fluctuations at the Jeans scale, both simulations having approximately the same.

Our use of the \cite{2000ApJ...530....1M} PDF in deriving $\Gamma$ is certainly an imperfect approximation.
In general, the details of the gas density PDF will depend on the exact cosmological model (in particular through the power spectrum of density fluctuations), the thermal state of the gas, as well as on the numerical properties of the simulation.
\cite{2005MNRAS.357.1178B}, in particular, studied the dependences of the inferred photoionization rate on these parameters, demonstrating that they have a significant effect and quantifying their importance.
Possibly of equal importance, however, is the entire thermal \emph{history} of the Universe.
In fact, the gas spatial distribution (through Jeans-like smoothing) depends not only on the present temperature of the gas, but also on its past state as it takes times for the gas to respond to changes \citep[e.g.,][]{1998MNRAS.296...44G}. 
In addition, fluctuations in the magnitude of the UV background and associated heating rates \citep[e.g.,][]{2004MNRAS.350.1107M}, the inhomogeneous reionization of both hydrogen and helium \cite[e.g.,][]{2003ApJ...596....9H, 2006ApJ...644...61L, 2007arXiv0711.0751F, 2007arXiv0711.1542F}, and secondary feedback effects from galaxy formation and evolution \cite[e.g.,][]{2006ApJ...638...52K} may also affect the detailed properties of the intergalactic gas.
A full investigation of these effects is beyond the scope of the present paper.

In what follows, we instead take a simple approach based on the \cite{2000ApJ...530....1M} PDF in deriving the hydrogen photoionization rate, focusing on its implications for the ionizing sources.
To estimate the effect of this assumed PDF, we have repeated our analysis with the gas density PDF from the Q5 hydrodynamical simulation of \cite{2003MNRAS.339..312S}.
This simulation of a $\Lambda$CDM universe with $(\Omega_{m},~\Omega_{\Lambda},~h,~\sigma_{8})=(0.3,~0.7,~0.7,~0.9)$ has a box side length of 10 $h^{-1}$ Mpc, with $324^{3}$ dark matter particles and as many baryonic ones.
This PDF yields a $\Gamma$ at $z=3$ higher than the one inferred using the \cite{2000ApJ...530....1M} by $\sim10\%$, all other parameters fixed.
While this is a non-negligible effect, it is comparable to the statistical error we quote (Table \ref{our measurement}) and it is relatively small in comparison with the total plausible systematic error, the magnitude of which can be estimated from the scatter between the existing measurements (Figure \ref{gamma comparison}).
We caution that more work is clearly needed to refine the accuracy of the measured $\Gamma$, but note that the conclusions that we develop here are broadly robust to an order unity uncertainty in its normalization that may reasonably be expected.
On the other hand, our results are consistently derived and finely binned over a large redshift range, clearly constraining the constancy of $\Gamma$ barring any strongly redshift-dependent systematic effect.

\begin{deluxetable}{lcccccl}
\tablewidth{0pc}
\tablecaption{\cite{2000ApJ...530....1M} Gas Density PDF Parameters\label{MHR params}}
\tabletypesize{\footnotesize}
\tablehead{\colhead{$z$} & \colhead{$A$\tablenotemark{a}} & \colhead{$\delta_{0}$} & \colhead{$b$} & \colhead{$C_{0}$}}
\startdata
2 & 0.176 & 2.54 & 2.23 & 0.558 \\
3 & 0.242 & 1.89 & 2.35 & 0.599 \\
4 & 0.309 & 1.53 & 2.48 & 0.611 \\
6 & 0.375 & 1.09 & 2.50 & 0.880 \\
\enddata
\tablenotetext{a}{The values given by \cite{2000ApJ...530....1M} are off by a factor of $\ln{10}$.}
\end{deluxetable} 

\subsection{Thermal History of the IGM}
Although the thermal history of the IGM may only have a modest effect on the gas density PDF, the parameters $T_{0}$ and $\beta$ must be known in order to unambiguously solve for the evolution of the background hydrogen photoionization rate, $\Gamma$.
In this paper, we do not attempt to infer $T_{0}$ and $\beta$ directly from the data; rather, we use existing measurements, which we combine with our theoretical understanding of the thermal physics of the IGM in order to assess the robustness of our conclusions.
We begin by reviewing this physics before examining the present observational constraints.

\subsubsection{Theory of the Thermal History of the IGM}
\label{thermal history theory}
A detailed formalism for the thermal history of the IGM can be found in \cite{1997MNRAS.292...27H}, or more concisely in the appendix of \cite{2003ApJ...596....9H}.
As we will ultimately base our analysis on empirical measurements, we simply summarize the most important physical processes here.

For given initial conditions, the temperature of a given fluid element is uniquely determined by a competition between cooling and heating processes.
In the low-density ($\delta\lesssim5$) IGM of the $z\lesssim10$ Universe, photoheating and adiabatic cooling typically drive the thermal evolution.
Photoheating proceeds in two main regimes.
During reionization events, the ionization fraction of a particular atomic species (with H and He being most important by a large factor) changes by a factor of order unity, inducing a large boost in the IGM temperature.
In an already ionized plasma, photoheating occurs through the ionization of small residual fractions of HI, HeI, or HeII.
In equilibrium, this process is balanced by recombinations.
Gas elements further heat or cool simply because of adiabatic contraction or expansion; the overall expansion of the Universe, for example, induces a temperature fall-off $T\propto(1+z)^{2}$.

Ultimately, the IGM temperature loses memory of reionization events, reaching a ``thermal asymptote'' determined by the balance between heating and cooling processes in the ionized plasma.
The thermal asymptote depends only on the shape of the background spectrum.
For a power-law ionizing background $J_{\nu}\propto \nu^{-\alpha}$ it is given by
\begin{align}
\label{thermal asymptote power law}
T_{0}=&2.49\times10^{4}{\rm~K}(0.417 + 0.047y_{\rm II} + 0.583y_{\rm III}) \\ \notag
& \times (2 + \alpha)^{-1/1.7}
\left(
\frac{1+z}{4.9}
\right)^{0.53}
\end{align}
\citep[][]{1997MNRAS.292...27H}.

What are relevant initial conditions?
Hydrogen reionization can heat the IGM by a few times $10^{4}$ K \citep[][]{1994MNRAS.266..343M, 1999ApJ...520L..13A, 2007MNRAS.380.1369T}, with negligible pre-reionization temperatures $\sim10^{3}$ K expected from heating by X-rays \citep[][]{2001ApJ...553..499O, 2001ApJ...563....1V, 2006ApJ...637L...1K, 2006MNRAS.371..867F}.
While we know for certain that the Universe is reionized by $z=6$ from quasar spectra \citep[e.g.,][]{2006ARA&A..44..415F}, the Thomson scattering optical depth measured by WMAP \citep[][]{2007ApJS..170..377S} indicates that it is likely to have proceeded at $z\lesssim20$.
Similar entropy injection may occur at lower redshifts ($z\sim3-4$) during the reionization of HeII \citep[e.g.,][]{2007arXiv0711.1542F}.

The parameters $(T_{0},~\beta)$ of the temperature-density relation (eq. \ref{temperature density relation}) can in principle be solved for by calculating the evolution of the temperature of fluid elements as a function of density.
Uncertainties in the redshifts of reionization events and their character, and on the evolution of the ionizing background spectrum however preclude a reliable \emph{ab initio} calculation of the temperature-density relation.
We therefore turn to empirical estimates.

\subsubsection{Thermal History Measurements}
The thermal evolution of the IGM can be probed using the \Lya~forest itself.
Indeed, the Jeans-like smoothing of the gas depends on its pressure, which in turn is set by its temperature.
In addition, \Lya~forest spectral features are thermally broadened.
Both of these effects increase the characteristic width of \Lya~absorption lines.

In general, Hubble broadening (arising from the differential Hubble flow across individual absorbers) and peculiar velocities also contribute to the line widths \citep[][]{1994ApJ...431..109M, 1997seim.proc..133W, 1999ApJ...517..541H, 2000MNRAS.315..600T}.
On ``velocity caustics,'' where the peculiar velocities of the gas particles cancel the Hubble gradient, however, the line width should be uniquely determined by the gas temperature.
This results in a cut-off in the distribution of line width versus column density that traces the temperature-density relation of the gas \citep[][]{1999MNRAS.310...57S, 2000MNRAS.318..817S, 2000ApJ...534...41R, 2001ApJ...562...52M}.

An alternative, but physically related, method to estimate the IGM temperature from the \Lya~forest is to consider its transmission small-scale power spectrum.
In this case, the finite temperature of the gas suppresses the small-scale power \cite[e.g.,][]{2000MNRAS.315..600T, 2000MNRAS.317..989T, 2001ApJ...557..519Z, 2002MNRAS.332..367T, 2002ApJ...564..153Z}.
\cite{2001ApJ...557..519Z} used this dependence of the small-scale power to jointly estimate $T_{0}$ and $\beta$ from the power spectrum measurement from eight Keck/HIRES spectra of \cite{2000ApJ...543....1M}.
The $1\sigma$ constraints on $T_{0}$ obtained after marginalizing over $\beta$ are $T_{0}=(2.10\pm0.45)\times10^{4}$ K at $z=2.4$, $T_{0}=(2.3\pm0.35)\times10^{4}$ K at $z=3$, and $T_{0}=(2.2\pm0.2)\times10^{4}$ K at $z=3.9$.\footnote{These values are given at $2\sigma$ in \cite{2003ApJ...596....9H}.}

We adopt the \cite{2001ApJ...557..519Z} values, consistent with those obtained by \cite{2001ApJ...562...52M} from the same data set but with an independent line-fitting method, in our analysis.
In the range $2.4\leq z\leq3.9$, we linearly interpolate between their $T_{0}$ measurements; for $z<2.4$, we linearly extrapolate based on the $z=2.4$ and $z=3$ values.
In our fiducial model, we assume that the IGM has been cooling adiabatically, $T_{0}\propto(1+z)^{2}$, for $z>3.9$.
For the slope of the temperature-density relation, we take the early reionization limit $\beta=0.62$.

The $1\sigma$ errors in our inferred $\Gamma$ (Figure \ref{gamma comparison} and Table \ref{our measurement}) include both the uncertainty in our $\tau_{\rm eff}$ measurement and on $T_{0}$.

\subsection{Ionization State of the IGM}
\label{ionization state}
We adopt the mass fractions $X=0.75$ and $Y=0.25$ for hydrogen and helium, respectively \citep[][]{2001ApJ...552L...1B}. 
At the redshifts $z<6$ of interest, quasar spectra indicate that the Universe is fully ionized, $x=1$, to better than one part in a thousand.
While helium is likely also almost fully ionized at $z\lesssim3$ ($y_{\rm II}=0$, $y_{\rm III}=1$), as indicated by observations of the HeII \Lya~forest, little is known about its ionization state at higher redshifts.
Owing to its similar ionization potential, helium is expected to be singly ionized simultaneously to hydrogen.
A number of observational lines of evidence \citep[for a review, see][]{2008ApJ...681..831F} as well as observationally calibrated theoretical models \citep[][]{2002MNRAS.332..601S, 2003ApJ...586..693W, 2007arXiv0711.1542F} suggest that HeII is reionized around $z=3-4$.
In our inference of $\Gamma$, we assume that helium is fully ionized at all epochs probed; $\Gamma$ may be overestimated by up to 8\% at the highest redshifts if our measurement reaches pre-HeII reionization times, owing to an overestimate of the number of free electrons (eq. \ref{n electrons}).

\subsection{Systematic Effects from the Thermal History}
\label{thermal history systematics}
We conclude this section by exploring systematic effects that may arise if our assumptions regarding the thermal history of the IGM are in error with respect to our derived photoionization rate.

\subsubsection{Temperature at Mean Density}
\begin{figure}[ht]
\begin{center}
\includegraphics[width=0.49\textwidth]{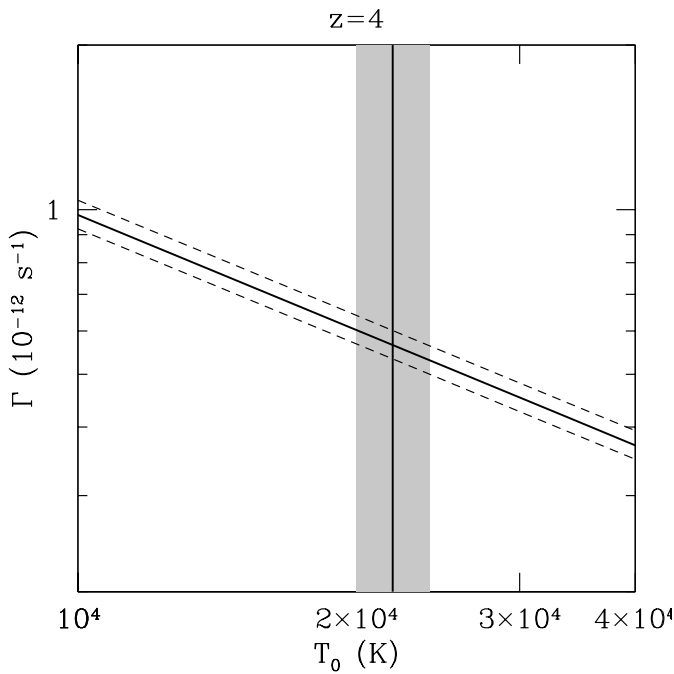}
\end{center}
\caption{Inferred hydrogen photoionization rate at $z=4$ as a function of the assumed temperature of the IGM at mean density.
The dashed lines show the statistical error in the derived $\Gamma$ from the error in the measured $\tau_{\rm eff}$ only.
The vertical line indicates the temperature at mean density $T_{0}=(2.2\pm0.2)\times10^{4}$ K measured by \cite{2001ApJ...557..519Z} at $z=3.9$.
The shaded area covers the 1$\sigma$ interval on this measurement.
All other parameters as in our fiducial inference.}
\label{gamma inferred varying t0}
\end{figure}

For fixed cosmology, ionization state of the IGM, and temperature-density relation slope $\beta$, $\tau_{\rm eff}$ constrains the quantity $T_{0}^{-0.7}/\Gamma$ (\S \ref{lya effective optical depth}).
An error in the temperature of the IGM at mean density, $T_{0}$, will therefore directly result in a corresponding error in the inferred $\Gamma$.
Although our error bars on $\Gamma$ include a contribution from the statistical uncertainty in the measured $T_{0}$, it is conceivable that these measurements suffer from systematic effects.
Figure \ref{gamma inferred varying t0} shows exactly how the inferred $\Gamma$ depends on the assumed $T_{0}$ at $z=4$.
We focus on this particular redshift as we will be interested in the robustness of a high $\Gamma$ at the highest redshifts probed when we consider the sources of the UV background (\S \ref{comparison with forest}).\\ \\

Note that the slight upward bump\footnote{The statistical significance of this feature is greater in the $\tau_{\rm eff}$ evolution than in the derived evolution of $\Gamma$, as the statistical errors in the latter include contribution from the error in $T_{0}$ which is correlated between bins. For an analysis of this feature in $\tau_{\rm eff}$, see~\cite{2008ApJ...681..831F}.} near $z=3.2$ in the derived $\Gamma$ (Fig. \ref{gamma comparison}) may be an artifact of our smooth interpolation of the temperature between the sparse measurements at $z\sim$ 2, 3, and 4.
This procedure directly translates a downward bump in $\tau_{\rm eff}$ into an upward bump in $\Gamma$.
An alternative possibility is that the evolution of $\Gamma$ is featureless and that $T_{0}$ instead undergoes a rapid increase, for example owing to the reionization of HeII \citep[theoretical work however suggests that the time scale for heating the IGM during HeII reionization is too long to explain such a narrow feature;][]{2008arXiv0807.2447B, 2008arXiv0807.2799M}.
For a discussion of these degenerate effects, see \cite{2008ApJ...681..831F}.
The detection of a bump in $\tau_{\rm eff}$ is more robust in that it is not subject to this degeneracy.

\subsubsection{Slope of the Temperature-Density Relation}
From line-fitting analyses, \cite{2000ApJ...534...41R} and \cite{2000MNRAS.318..817S} have found some (albeit marginal) evidence that the temperature-density relation becomes isothermal near $z\approx3$, perhaps owing to reheating from the reionization of HeII near this redshift.
Comparing an improved measurement of the \Lya~forest flux PDF at $1.7<z<3.2$ by \cite{2007arXiv0711.1862K} to a large set of hydrodynamical simulations with different cosmological parameters and thermal histories, \cite{2007arXiv0711.2064B} find evidence for an equation that is close to isothermal, or even inverted ($\beta<0$).
From a data set extending to $z=5.8$, \cite{2007ApJ...662...72B} also find suggestive evidence that an inverted temperature-density relation with $\beta\approx-0.5$ fits the observed \Lya~flux PDF best, if the underlying gas density PDF follows that of \cite{2000ApJ...530....1M}.
This suggests that the voids in the IGM may be significantly hotter and the thermal state of the low-density IGM more complex than previously assumed.
These claims have not been corroborated by the results of \cite{2001ApJ...562...52M} and \cite{2001ApJ...557..519Z}, although both these analyses are consistent with isothermality at $z\leq3$.

While the present observational evidence is scarce and difficult to interpret, some theoretical modeling of inhomogeneous HeII reionization also suggests a complex, possibly inverted temperature-density relation \citep[][]{2004MNRAS.348L..43B, 2005MNRAS.361.1399G, 2007MNRAS.380.1369T, 2007arXiv0711.0751F}.
Detailed radiative transfer simulations of HeII reionization however find that the slope of the temperature-density relation is only moderately affected and that in actuality the correct slope is somewhere between the early hydrogen reionization limit ($\beta=0.62$) and the isothermal case ($\beta=0$) \citep[][]{2008arXiv0807.2799M}.
These simulations also produce a large scatter in the temperature-density relation, suggesting that a power-law approximation is rather rough during HeII reionization.

\begin{figure}[ht]
\begin{center}
\includegraphics[width=0.49\textwidth]{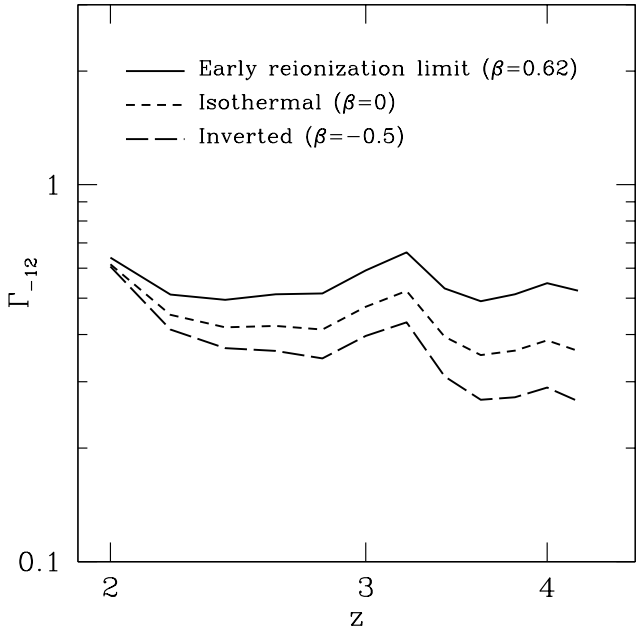}
\end{center}
\caption{Dependence of the inferred hydrogen photoionization rate on the low-density IGM temperature-density relation slope $\beta=\gamma-1$ (eq. \ref{temperature density relation}).
We show the results for the early reionization limit, the isothermal case, and an inverted relation.
As the temperature-density relation becomes flatter or inverted, the inferred $\Gamma$ at $z\geq2$ decreases, indicating that the \Lya~forest becomes more sensitive to underdense regions with $\delta<0$.
This is increasingly the case at higher redshifts.
Near $z=2$, the inferred $\Gamma$ is nearly independent of the slope of the temperature-density relation, while at $z=4$ the early-reionization limit gives a $\Gamma$ that is about twice that of the inverted case.
All other parameters as in our fiducial inference.}
\label{gamma inferred varying beta}
\end{figure}

How does the uncertainty in the slope of the temperature-density relation of the low-density IGM affect our inference of $\Gamma$?
In Figure \ref{gamma inferred varying beta}, we repeat our calculation of $\Gamma$ as determined from our effective optical depth measurement varying $\beta$.
We consider, in addition to the early hydrogen reionization limit, isothermal and inverted ($\beta=-0.5$) temperature-density relations.
The latter is consistent with the best-fit parameters found by \cite{2007arXiv0711.2064B} and \cite{2007ApJ...662...72B}, but likely too extreme to be physically realistic based on the calculations of \cite{2008arXiv0807.2799M}.

As the temperature-density relation becomes flatter or inverted, the inferred $\Gamma$ at $z\geq2$ decreases, indicating that the \Lya~forest becomes more sensitive to underdense regions with $\delta<0$.
This is increasingly the case at higher redshifts.
Near $z=2$, the inferred $\Gamma$ is nearly independent of the slope of the temperature-density relation (indicating that the measurement is most sensitive to $\delta \approx0$), while at $z=4$ the early-reionization limit gives a $\Gamma$ that is about twice that of the inverted case.

Note that the systematic uncertainties in the temperature of the IGM at the mean density $T_{0}$ and on the slope of the temperature-density relation $\beta$, especially if they conspire to affect the inferred $\Gamma$ in the same direction, could go some way in resulting in a declining photoionization rate toward high redshifts $z\gtrsim3$ and therefore bringing it in better agreement with observationally derived star formation rates (see \S \ref{comparison with forest}).
However, the requirement that the Universe be reionized by $z=6$ also argues forcefully against a steep decline of star formation beyond $z=3$ (\S \ref{reionization constraints}).
The latter argument, depending only on the magnitude of $\Gamma$ at $z=3$ through the normalization of the ionizing emissivity, is largely insensitive to the uncertainties in the thermal history at the redshifts probed by our \Lya~forest opacity measurement.

\section{FROM PHOTOIONIZATION TO EMISSIVITY}
\label{photoionizations to emissivity}
Although $\Gamma$ is clearly related to the cosmic sources of ionizing photons, the sources are most closely related to their specific emissivity, $\epsilon_{\nu}$.
In this section, we compute the UV emissivity implied by our inferred photoionization rate, before proceeding to investigate its implications for quasars and star-forming galaxies in the next.
Our formalism is based on \cite{1996ApJ...461...20H}, \cite{1999ApJ...514..648M}, and \cite{2003ApJ...584..110S}.

\subsection{Radiative Transfer}
\label{radiative transfer}
This specific intensity of the UV background obeys the cosmological radiative transfer equation,
\begin{equation}
\label{transfer equation}
\left(
\frac{\partial}{\partial t} - \nu H \frac{\partial}{\partial \nu}
\right)
J_{\nu}
=
-3 H J_{\nu}
-c \alpha_{\nu} J_{\nu}
+ \frac{c}{4\pi}\epsilon_{\nu},
\end{equation}
where $H(t)$ is the Hubble parameter, $c$ is the speed of light, $\alpha_{\nu}$ is the proper absorption coefficient, and $\epsilon_{\nu}$ is the proper emissivity.
Integrating equation (\ref{transfer equation}) and expressing the result in terms of redshift gives
\begin{equation}
\label{transfer equation solution}
J_{\nu_{0}}(z_{0})
=\frac{1}{4 \pi}
\int_{z_{0}}^{\infty}
dz
\frac{dl}{dz}
\frac{(1+z_{0})^{3}}{(1+z)^{3}} \epsilon_{\nu}(z) \exp[-\bar{\tau}(\nu_0, z_{0}, z)],
\end{equation}
where $\nu=\nu_{0}(1+z)/(1+z_{0})$, the proper line element $dl/dz=c/[(1+z)H(z)]$, and $\bar{\tau}$ quantifies the attenuation of photons of frequency $\nu_{0}$ at redshift $z_{0}$ that were emitted at redshift $z$.
For Poisson-distributed absorbers, each of column density $N_{\rm HI}$,
\begin{equation}
\label{taueff poisson expression}
\bar{\tau}(\nu_{0}, z_{0}, z) =
\int_{z_{0}}^{z} dz'
\int_{0}^{\infty}
dN_{\rm HI}
\frac{\partial^{2}N}{\partial N_{\rm HI} \partial z'}
(1 - e^{-\tau_{\nu}}),
\end{equation}
where $\partial^{2}N/\partial N_{\rm HI} \partial z'$ is the column density distribution versus redshift and $\tau_{\nu}=N_{\rm HI}\sigma_{\rm HI}(\nu)$ \citep{1980ApJ...240..387P}.
We neglect the continuum opacity owing to helium and heavier elements, which is a good approximation near the hydrogen Lyman edge. 
For the redshifts $z\gtrsim2$ of interest here, the mean free path $\Delta l(\nu)$ of UV ionizing photons is so short that the sources contributing to $\Gamma$ at a given point are largely local.
In this local-source approximation \citep[e.g.,][]{2003ApJ...584..110S}, the solution to equation (\ref{transfer equation solution}) reduces to
\begin{equation}
\label{local source approximation}
J_{\nu}(z)\approx
\frac{1}{4\pi}
\Delta l(\nu,~z) \epsilon_{\nu}(z). 
\end{equation}
We may thus trivially solve for the emissivity,
\begin{equation}
\label{local source approximation emissivity}
\epsilon_{\nu}(z)
\approx
4\pi 
\frac{J_{\nu}(z)}{\Delta l(\nu,~z)}.
\end{equation}

\subsection{Mean Free Path}
\label{mean free path section}
As the mean free path of ionizing photons is the key quantity relating $\Gamma$ to its sources, we pause to derive an improved value in the light of new data and estimate its uncertainty.
We assume that the column density distribution of absorbers is well-approximated by power laws of the form
\begin{equation}
\label{joint NHI z distribution}
\frac{\partial^{2}N}{\partial N_{\rm HI} \partial z}
= N_{0} N_{\rm HI}^{-\beta} (1+z)^{\gamma},
\end{equation}
in which case the general solution to equation (\ref{taueff poisson expression}) is
%\begin{equation}
\begin{align}
\label{taueff poisson solution}
\bar{\tau}(\nu_{0},~z_{0},~z)=
\frac{\Gamma(2-\beta)}{(\beta-1)(\gamma-3\beta+4)} \notag \\
\times N_{0} \left(\frac{\nu_{\rm HI}}{\nu_{0}} \right)^{3(\beta-1)}
\sigma_{\rm HI}^{\beta-1}
(1+z_{0})^{3(\beta-1)} \notag \\
\times \left[
(1+z)^{\gamma-3\beta+4} - (1+z_{0})^{\gamma-3\beta+4}
\right].
\end{align}
%\end{equation}
Here, $\Gamma(2-\beta)$ is the gamma-function and is not related to the photoionization rate.
This reduces to the expression given by \cite{1999ApJ...514..648M} in the special case $(\beta,~\gamma)=(1.5,~2)$.
The mean free path is obtained by differentiating, setting $\Delta\bar{\tau}=1$, and using the line element $dl/dz$ to convert to proper length:
\begin{align}
\label{mean free path equation}
\Delta l(\nu_{0},~z)\approx
\frac{(\beta-1)c}{\Gamma(2-\beta)N_{0}\sigma_{\rm HI}^{\beta-1}} 
\left(\frac{\nu_{0}}{\nu_{\rm HI}}\right)^{3(\beta-1)}
\notag \\
\times
\frac{1}
{(1+z)^{\gamma+1}H(z)}.
\end{align}

For steep column density distributions with $\beta \approx 1.5$, as suggested by observations, most of the contribution to $\bar{\tau}$ at the Lyman limit arises for systems of optical depth near unity ($N_{\rm HI}\approx10^{17.2}$ cm$^{-2}$).
We thus focus on the values of the power-law indexes $\beta$ and $\gamma$ that are most appropriate in this neighborhood.
\cite{1995ApJ...444...64S} find that $dN/dz=C(1+z)^{\gamma}$ with $C=0.25^{+0.17}_{-0.10}$ and $\gamma=1.50\pm0.39$ for systems with $N_{\rm HI}\geq10^{17.2}$ cm$^{-2}$, whereas the column density distribution is well-fitted
by $\beta=1.390\pm0.027$ \citep[][]{2007AJ....134.1634M}.
The constant $N_{0}$ in equation (\ref{joint NHI z distribution}) is related to $C$ by $N_{0}=(\beta-1)C N_{\rm HI,min}^{\beta-1}$, where $N_{\rm HI,min}=10^{17.2}$ cm$^{-2}$ is the minimum column density accounted for in $dN/dz$.

\begin{figure}[ht]
\begin{center}
\includegraphics[width=0.49\textwidth]{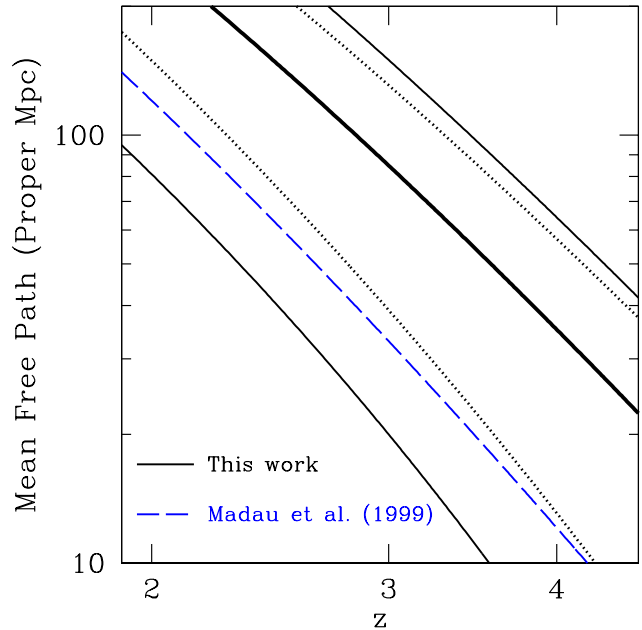}
\end{center}
\caption{Mean free path of photons at the Lyman limit versus redshift.
The thick solid curve shows our calculation based on the column density distribution of absorbers measured by \cite{2007AJ....134.1634M} and the redshift number density reported by \cite{1995ApJ...444...64S}.
The bounding thinner solid curves indicate the propagated uncertainty in the mean free path.
The dotted curves show the errors taking into account only the uncertainty in the redshift evolution of the mean free path ($\gamma$).
The blue long-dashed curve shows the commonly used mean free path of \cite{1999ApJ...514..648M}.}
\label{mfp figure}
\end{figure}

Figure \ref{mfp figure} shows the corresponding mean free path at the Lyman limit, $\Delta l\approx85[(1+z)/4]^{-4}$ proper Mpc, and its propagated uncertainty.
The smaller errors take into account only the uncertainty in the redshift evolution of the mean free path ($\gamma$).
Also shown is the commonly used result of \cite{1999ApJ...514..648M}, $\Delta l^{\rm Madau}\approx33[(1+z)/4]^{-4.5}$ proper Mpc, which is seen to underestimate our value by a factor $\approx2.5$, albeit within the uncertainty.
The redshift evolution differs by a power of 0.5 as we have adopted the power-law index $\beta=1.5$ from \cite{1995ApJ...444...64S} for the redshift evolution of Lyman limit systems, consistent with the results of 
\cite{1994ApJ...427L..13S}.

\cite{1999ApJ...514..648M} motivate their steeper choice of $\gamma=2$ from measurements of the redshift evolution of optically thin systems of lower column density \citep[][]{1993ApJ...418..585P, 1994ApJS...91....1B, 1997AJ....114....1K}.
However, if, for example, Lyman limit systems are associated with the outskirts of galaxies \citep[e.g.,][]{1982Natur.298..427T, 1984ApJ...281...76B, 1996ApJ...457L..57K, 1988ApJ...332...96L, 1989ApJS...69..703S, 1990ApJS...74...37S, 2004AJ....128.2954M, 2007ApJ...655..685K}, then their evolution may not follow that of absorbers in the diffuse IGM.
We therefore prefer to use the column density distribution measurements directly probing Lyman limit systems.
Although it is true that lower column density systems contribute 
non-negligibly to the Lyman continuum opacity, we note that instead choosing $\gamma=2$ would lead to a steeper decrease of the mean free path with increasing redshift.
This in turn would result in a larger requirement on the ionizing emissivity in order to produce the observed photoionization rate, exacerbating the tension we find in \S \ref{comparison with forest} and \S \ref{sfr results} with some estimates of the star formation rate density at high redshifts.
In this sense, our choice is conservative and our conclusions would only be strengthened if we had significantly underestimated the redshift evolution of the absorption systems determining the mean free path.

We finally note that even for identical choices of ($N_{0}$,~$\beta$,~$\gamma$), our estimate of the mean free path would exceed that of \cite{1999ApJ...514..648M} by about 30\% owing to their assumption of an Einstein-de Sitter cosmology, in contrast to the \emph{WMAP} cosmology we adopt. 

\subsection{Emissivity}
\label{emissivity}
We finally turn to the calculation of the emissivity at the Lyman limit implied by our measurement of $\Gamma$.
Assuming that the ionizing background has a power-law spectrum $J_{\nu}=J_{\nu_{\rm HI}}(\nu/\nu_{\rm HI})^{-\alpha_{\rm HI}}$ blueward of the Lyman limit, 
\begin{equation}
J_{\nu_{\rm HI}} =
\frac{\Gamma h (\alpha_{\rm HI}+3)}{4\pi \sigma_{\rm HI}},
\end{equation}
from which
\begin{equation}
\label{epsilon LL from Gamma}
\epsilon_{\nu_{\rm HI}}(z)\approx
\frac{\Gamma(z) h (\alpha_{\rm HI}+3)}{\sigma_{\rm HI} \Delta l(\nu_{\rm HI},~z)},
\end{equation}
or, equivalently,
\begin{equation}
\label{Gamma from epsilon LL}
\Gamma(z) \approx
\frac{\sigma_{\rm HI} \Delta l(\nu_{\rm HI},~z)  \epsilon_{\nu_{\rm HI}}(z)}{h (\alpha_{\rm HI}+3)} \, .
\end{equation}

We are also interested in the rest-frame UV emissivity probed by optical quasar and galaxy surveys.
Following the galaxy survey literature, we define the UV emissivity $\epsilon_{\rm UV}$ to correspond to the wavelength 1500~\AA, and denote the corresponding frequency by $\nu_{\rm UV}$.
If $\alpha_{\rm UV}$ is the spectral index between the Lyman limit and 1500~\AA, then
\begin{equation}
\label{epsilon UV from epsilon LL}
\epsilon_{\rm UV}=
\frac{1}{f_{\rm esc}}
\left(
\frac{1500~{\rm \AA}}{912~{\rm \AA}}
\right)^{\alpha_{\rm UV}}
\epsilon_{\nu_{\rm HI}}.
\end{equation}
The escape fraction $f_{\rm esc}$ accounts for the fact that, owing to Lyman continuum opacity associated with the host halo, only a certain fraction of the photons blueward of the Lyman limit escape from their source.
The correct value depends on the nature of the sources, to which we now turn.

\section{THE COSMIC SOURCES OF UV PHOTONS}
\label{cosmic sources}
The two known dominant sources of UV photons in the Universe are quasars and star-forming galaxies.
We investigate the constraints that can be put on their populations using our measurement of the IGM \Lya~opacity.

Our starting assumption is that the luminosity function of quasars and their spectral energy distribution (SED) are better constrained than those of galaxies.
In addition, we assume that all the photons capable of ionizing hydrogen emitted by quasars escape their host unimpeded, i.e. that quasars have an escape fraction of unity.
This is often assumed to be the case on the theoretical basis that quasars are sufficiently powerful to fully ionize their surrounding gas \citep[e.g.,][]{2000ApJ...545...86W}.
Observational evidence that this is the case is provided by the proximity effect, which is clearly detected at roughly the level predicted if the SED of quasars does not suffer a break at the Lyman limit from intrinsic absorption \citep[e.g.,][]{2000ApJS..130...67S, 2007arXiv0711.4113G, 2008arXiv0801.1767D}.
Moreover, quasar spectra generally do not show DLAs associated with the quasar hosts.
In contrast, DLAs of high column density are seen in most spectra of long-duration gamma-ray bursts occurring in star-forming galaxies and their Lyman-continuum opacity is understood to be responsible for the small escape fraction of inactive galaxies \citep[][]{2007ApJ...667L.125C, 2007arXiv0707.0879G}.

\subsection{Quasar Emissivity}
\label{quasar emissivity}
\cite{2007ApJ...654..731H} have used a large set of observed quasar luminosity functions in the infrared, optical, soft and hard X-rays, as well as emission line measurements, combined with recent estimates of the quasar column density distribution from hard X-ray and IR observations and measurements of their spectral shape from the radio through hard X-rays to estimate their bolometric luminosity function.
Denoting by $\epsilon_{B}$ the emissivity at 4400~\AA~and assuming $L_{B}\equiv \nu L_{\nu}|_{4400~{\rm \AA}}$,
\begin{equation}
\label{quasar B emissivity}
\epsilon_{B}^{\rm com} =
\int_{0}^{\infty} dL_{B} 
\frac{d\phi}{dL_{B}}
\frac{L_{B}}{\nu|_{\rm 4400~\AA}},
\end{equation}
where $d\phi/dL_{B}$ is the $B$-band luminosity function derived by \cite{2007ApJ...654..731H} from their full spectral and obscuration modeling in comoving units.
To convert to an emissivity at the Lyman limit, we assume that quasars have a spectral index $\alpha=0.3$ at 2500-4400~\AA, 0.8 at 1050-2500~\AA~\citep[][]{1999ApJ...514..648M}, and 1.6 from 1050~\AA~to 228~\AA~\citep[][]{2002ApJ...565..773T}.
The corresponding photoionization rate is then obtained using equation (\ref{epsilon LL from Gamma}).

Although equation \ref{quasar B emissivity} appears to assume isotropic quasar emissivity, it is more generally valid.
For instance, if quasars emit with constant intensity in some directions and are completely obscured toward the others, then the overestimate of the total photon output of a quasar based on its observed magnitude and the assumption of isotropy will be exactly canceled by the fraction of quasars missed in the luminosity function.
To generalize further, note that in a context in which quasars shine with different intensities in different directions, the observed quasar luminosity function can be interpreted as quantifying the number of (randomly oriented) lines of sights of given intensity from quasars.
For the purpose of calculating the cosmological emissivity, one can imagine distributing these sight lines among isotropically emitting sources and recover equation \ref{quasar B emissivity}.
 
Some biases could however be introduced in the calculation if the angular distribution of the quasar intensity has a more complex smoothly varying profile or depends on frequency (see \S \ref{comparison with forest}).

We take a background spectral index shortward of the Lyman limit $\alpha_{\rm HI}=0.1$, appropriate if the spectrum is hardened by IGM filtering.
This hardening arises from the frequency dependence of the mean free path of ionizing photons (eq. \ref{mean free path equation}) and the relation between the UV background spectrum and the emissivity producing it (eq. \ref{local source approximation emissivity}): $J_{\nu}\propto \epsilon_{\nu} \nu^{3(\beta-1)}$.
Taking the approximate value $\beta = 1.5$ for the column density distribution of absorbers, the spectrum is hardened by $\alpha \rightarrow \alpha-1.5$.\footnote{Note that this differs from the maximum hardening $\alpha \rightarrow \alpha-3$ that would result from a homogeneous distribution of neutral hydrogen because of the $\sigma(\nu)\propto \nu^{-3}$ photoionization cross-section dependence.
The difference arises from the distribution of absorbers into discrete clumps and depends on the slope of the column density distribution.
See, for example, the argument in \cite{1993ApJ...418...28Z}.}

\subsection{Stellar Emissivity}
\label{stellar emissivity section}
Because only massive short-lived stars produce UV photons in significant amounts, the stellar emissivity is intimately connected to the cosmic SFR.
To calculate the stellar contribution to the emissivity of ionizing photons, we consider both theoretical \citep[][]{2003MNRAS.339..312S, 2003MNRAS.341.1253H} and observational estimates \citep[][and references therein]{2004ApJ...615..209H, 2006ApJ...651..142H} of the cosmic star formation history, which we describe in \S \ref{sfr}.

The exact calculation of the emergent emissivity of ionizing photons from galaxies involves detailed assumptions about the initial mass function (IMF) of stars, stellar astrophysics, and the escape fraction of ionizing radiation, all of which are subject to substantial uncertainty.
Under the assumption that these are constant across cosmic time, however, the emissivity is simply proportional to the SFR density,
\begin{equation}
\label{emissivity propto sfr}
\epsilon_{\nu}=K\dot{\rho}_{\star}(1+z)^{3},
\end{equation}
where the factor of $(1+z)^{3}$ arises from the conversion from comoving SFR density to proper emissivity in our notation, and the constant $K$ depends on the uncertain astrophysics.
The usefulness of this description arises from the fact that significant insight can be gained from the redshift evolution of the emissivity only.
If the proper mean free path evolves as $\Delta l\propto(1+z)^{-4}$ (appropriate if the line-of-sight number density of Lyman limit systems evolves as $\partial N/\partial z\propto(1+z)^{1.5}$; \S \ref{mean free path section}), then we can use the local-source approximation in equation (\ref{local source approximation}) to relate the evolution of the comoving SFR density to that of the hydrogen photoionization rate:
\begin{equation}
\label{Gamma from sfr}
\Gamma = K' \dot{\rho}_{\star} (1+z)^{-1},
\end{equation}
for some other constant $K'$.

\subsubsection{Star Formation Rate Density}
\label{sfr}
Employing hydrodynamic simulations of structure formation in a $\Lambda$CDM cosmology, \cite{2003MNRAS.339..312S} calculated the comoving SFR density $\dot{\rho}_{\star}(z)$, taking into account radiative heating and cooling of the gas, star formation, supernova feedback, and galactic winds based on the hybrid star-formation multiphase model detailed in \cite{2003MNRAS.339..289S}.
They obtained a SFR density gradually rising by a factor of ten from $z=0$ and peaking at $z\sim5-6$.
Using simple analytic reasoning, \cite{2003MNRAS.341.1253H} identified the processes that drive the evolution of the cosmic SFR in CDM universes, and derived an analytic model matching their simulation results to better than $\approx10$\%.

They found that the cosmic SFR is described by two regimes.
At early times, densities are sufficiently high and cooling times sufficiently short that abundant quantities of star-forming gas are present in all dark matter halos that can cool by atomic processes.
In this gravitationally-dominated regime, $\dot{\rho}_{\star}$ generically rises exponentially as $z$ decreases, independent of the details of the physical model for star formation, but dependent on the normalization and shape of the cosmological power spectrum.
At low redshifts, densities decline as the Universe expands to the point that cooling is inhibited, limiting the amount of star-forming gas available.
In this regime, the SFR scales in proportion to the cooling rate within halos, $\dot{\rho}_{\star}\propto H(z)^{4/3}$.

These two regimes lead to a redshift of peak star formation which in general depends on the efficiency of star formation.
For a star formation efficiency normalized to the empirical Kennicutt law \citep[][]{1989ApJ...344..685K, 1998ApJ...498..541K} of disk galaxies at $z=0$, the comoving SFR density in their simulations is well described by the analytic form
\begin{equation}
\label{hernquist springel model}
\dot{\rho}_{\star}(z)=\dot{\rho}_{\star}(0)
\frac{\chi^{2}}{1+\alpha(\chi-1)^{3}\exp{(\beta \chi^{7/4})}},
\end{equation}
where
\begin{equation}
\chi\equiv
\left(
\frac{H(z)}{H_{0}}
\right)^{2/3}
\end{equation}
and $\alpha=0.012$, $\beta=0.041$, and $\rho_{\star}(0)=0.013$ M$_{\odot}$ yr$^{-1}$ Mpc$^{-3}$ are fitting parameters.

This fit was derived for their $\Lambda$CDM cosmology with $\Omega_{\rm m}=0.3$, $\Omega_{\Lambda}=0.7$, $h=0.7$, and $\sigma_{8}=0.9$.
For consistency, in this paper we use a modification of this fit adapted to the $WMAP$ cosmology that we assume.
Specifically, we use equation (45) of \cite{2003MNRAS.341.1253H} to evaluate their analytic model for the $WMAP$ cosmology.
The resulting star formation history differs from their fiducial $\Lambda$CDM model through the parameter $\sigma_{8}$ governing the amplitudes of density fluctuations.
Specifically, star formation is slightly delayed for $\sigma_{8}=0.82$ with respect to the higher value $\sigma_{8}=0.9$.

Extending prior work by \cite{2004ApJ...615..209H}, \cite{2006ApJ...651..142H} examined the observational constraints on the cosmic star formation history based on a database of measurements in all relevant wavebands.
To the luminosity density measurements, dust corrections (where necessary), SFR calibrations, and IMF assumptions are applied to derive the star formation history.
These authors chose to use the parametric form of \cite{2001MNRAS.326..255C} for their fits to the star formation history:
\begin{equation}
\label{cole sfr form}
\dot{\rho}_{\star}(z)=
\frac{(a+bz)h}{1+(z/c)^{d}},
\end{equation}
where $h=0.7$.
For a ``modified Salpeter A'' IMF \citep[][]{2003ApJ...593..258B}, they find the best-fit parameters $(a,~b,~c,~d)=(0.0170,~0.13,~3.3,~5.3)$.
Other IMFs, such as the one advocated by \cite{2003ApJ...593..258B}, affect this fit only in its overall normalization, provided they are constant with cosmic time.
The IMF normalization will be irrelevant for the following arguments and we therefore only consider the modified Salpeter A case.

Although we will focus on the widely used \cite{2006ApJ...651..142H} fit for convenience in our study, we note that it is representative of many other current state-of-the-art observational estimates of the SFR density in that it peaks at $z=2-3$ and subsequently declines \citep[e.g.,][]{2008ApJS..175...48R}.
In addition to the smooth fit based on the \cite{2001MNRAS.326..255C} parametric form, \cite{2006ApJ...651..142H} also present a piecewise linear fit of the same data in order to avoid artifacts resulting from the choice of a particular functional form.
We have duplicated our analysis with their piecewise linear fit as well and reached practically identical conclusions.
To simplify our presentation, we only show our calculations for the smooth fit.

\begin{figure*}[ht]
\begin{center}
\includegraphics[width=1.0\textwidth]{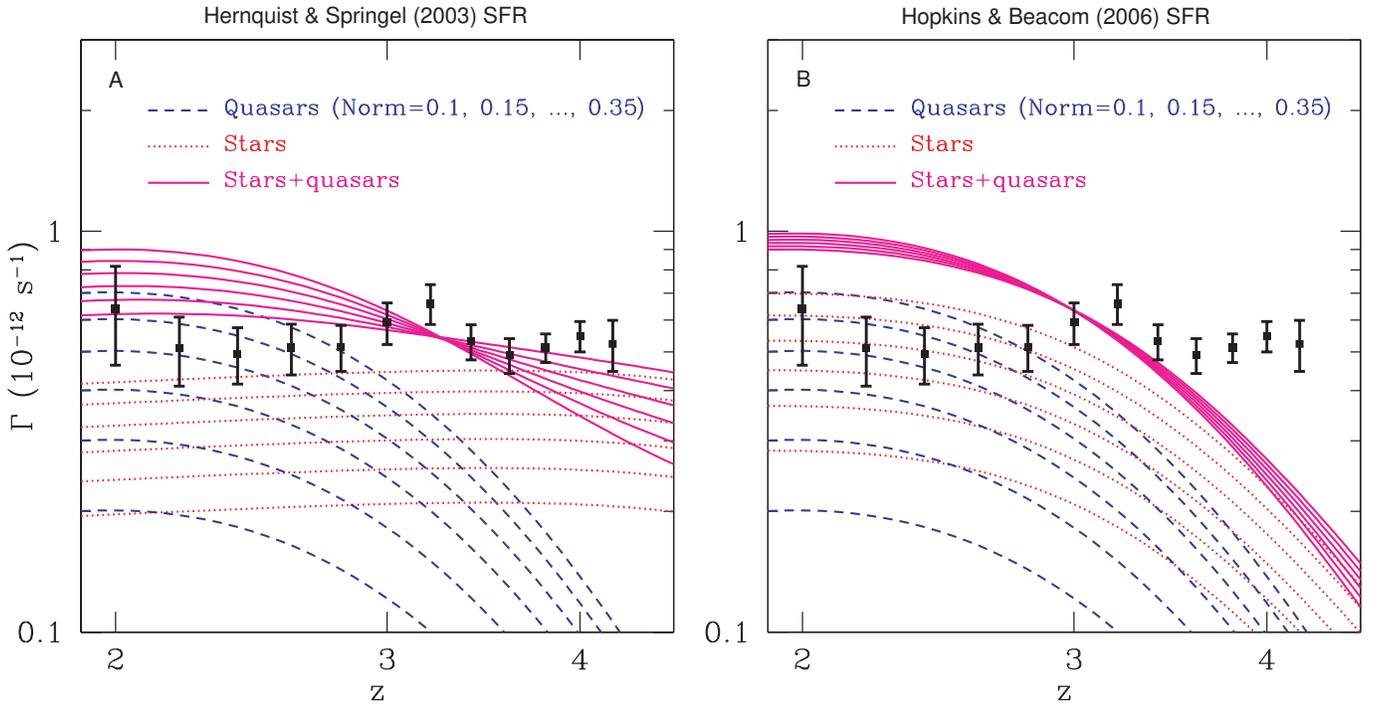}
\end{center}
\caption{Best-fit stellar (red dotted) and quasar+stellar (purple solid) contributions to the hydrogen photoionization rate for varying normalization of the quasar component calculated based on the luminosity function of \cite{2007ApJ...654..731H} (blue long-dashed).
The stellar ionizing emissivity is assumed to trace the star formation rate density in the theoretical model of \cite{2003MNRAS.341.1253H} (panel A) or the best-fit to observational estimates of \cite{2006ApJ...651..142H} (panel B).
In each case, the best-fit stellar normalization is calculated for a fixed quasar normalization, from 0.1 to 0.35, modeling the effects of an escape fraction less than unity or of an uncertainty in the normalization of the mean free path of ionizing photons.
Because the stellar contribution peaks near $z\approx2$ similarly to quasars for the \cite{2006ApJ...651..142H} star formation history, the sum of the two fails to account for the flat photoionization rate measured from the \Lya~forest over the redshift range $2\leq z \leq 4.2$.}
\label{best fit stars}
\end{figure*}

\subsection{Comparison with the \Lya~Forest Photoionization Rate}
\label{comparison with forest}
To compare the photoionization rate measured from the \Lya~forest to what can be accounted for by the observed quasars and estimated star formation history, we proceed as follows.
We begin by fixing the quasar contribution, with the working assumption that it is better constrained than that of stars.
For each star formation history we consider, we then solve for the constant $K'$ in equation (\ref{Gamma from sfr}) that minimizes the $\chi^{2}$ between our measured $\Gamma$ and the sum of the quasar and stellar contributions.
We repeat the exercise for different normalizations of the quasar contribution to investigate how our conclusions would be affected if the escape fraction of quasars were less than unity, or if there were an error in the normalization of the mean free path or in the assumed spectral indexes.

The results are shown in Figure \ref{best fit stars}.
The superpositions of the best-fit nearly flat contribution to $\Gamma$ from stars in the \cite{2003MNRAS.341.1253H} model and of the quasar contribution peaked near $z\approx2$ estimated from the \cite{2007ApJ...654..731H} luminosity function match the observed total photoionization rate from the \Lya~forest reasonably well (panel A).
However, if the stellar ionizing emissivity instead traces the fit of \cite{2006ApJ...651..142H} to the SFR density, then the stellar contribution to $\Gamma$ peaks similarly to that of quasars near $z\approx2$.
As a result, any superposition of the two contributions also peaks near $z\approx2$.
As the measured total $\Gamma$ is nearly flat between $z=2$ and $z=4.2$, any such superposition falls short of accounting for it at $z=4.2$ by a factor $\approx4$ while simultaneously overestimating its value by a factor $\approx2$ at $z=2$ (panel B), in the best-fit case.
Note that if we had assumed an abundance of Lyman-limit systems increasing with redshift as $(1+z)^{2}$ \citep[as in][and appropriate for lower column density systems]{1999ApJ...514..648M} instead of $(1+z)^{1.5}$, then the discrepancy at $z=4.2$ would be even larger.
Unless the escape fraction of star-forming galaxies or the IMF evolve appropriately with redshift, our measurement of the total photoionization rate from the \Lya~forest is therefore in conflict with the best-fit SFR density of \cite{2006ApJ...651..142H}.
We will discuss potential issues with this SFR estimate in \S \ref{madau diagram}.

Note that in Figure \ref{best fit stars} we have only shown quasar normalizations from 0.1 to 0.35, as for higher normalizations quasars alone overproduce the measured total photoionization rate, which is of course unphysical.
The fact that a normalization strictly less than unity is required may indicate that the photoionization rate derived from the forest is too low either because of an overestimate of the mean free path of ionizing photons or inexact spectral assumptions.
A smaller mean free path would yield, for fixed emissivity, a lower photoionization rate (eq. \ref{Gamma from epsilon LL}).
The ionizing emissivity, in turn, is obtained from modeling the quasar spectrum between the $B$-band at $\sim4400$~\AA~and the Lyman edge at 912~\AA.
The photoionization rate moreover depends on the spectral shape shortward of 912~\AA~and therefore some level of uncertainty in the calculated value definitely arises from the assumed spectrum.

It could also indicate that the escape fraction of Lyman continuum photons from quasars is in fact less than unity.
It is interesting to consider why this may be the case.
Although (unlike GRBs occurring in star-forming galaxies) quasar hosts typically do not show DLAs, Lyman limit systems are observed to cluster around quasars, apparently preferentially in the transverse direction \citep[][]{2006ApJ...651...61H, 2007ApJ...655..735H, 2008arXiv0806.0862P}.
It is possible that such systems absorb a significant fraction of the ionizing radiation emitted by the quasars they surround before these penetrate into the diffuse IGM, resulting into an ``effective'' escape fraction less than unity.
A similar effect may also be in action around star-forming galaxies, although Lyman limit systems may cluster more weakly around the latter owing to their lower typical masses.
As our argument here is based on the shape of the evolution of the photoionization rate, it is largely robust to the normalization uncertainties.

In any case, the apparent overproduction of the ionizing background by quasars alone highlights the uncertainties in converting from a luminosity function to a photoionization rate.
In particular, models that are based on integrating the luminosity functions of the ionizing sources \cite[e.g.,][]{1996ApJ...461...20H} risk being significantly in error in normalization (see \S \ref{haardt and madau}).
Calibrating these models to the photoionization rate measured from the \Lya~forest using the flux decrement method, while itself subject to some systematic effects (\S \ref{gamma}), should yield more robust results.

We now proceed to consider further constraints on the cosmic ultraviolet luminosity density imposed by the reionization of hydrogen and helium.

\begin{figure}[ht]
\begin{center}
\includegraphics[width=0.49\textwidth]{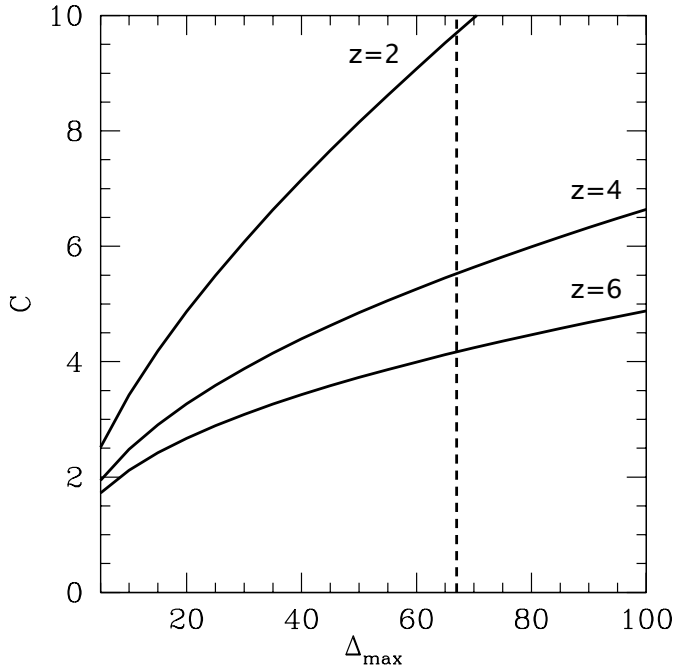}
\end{center}
\caption{Clumping factor calculated from the \cite{2000ApJ...530....1M} gas density PDF, $C\equiv \langle \Delta^{2}\rangle/\langle \Delta \rangle^{2}$, as a function of the maximum dimensionless overdensity $\Delta_{\rm max}$ included in the integration.
The curves are for $z=$2,~4,~and 6 from most to least clumpy.
The vertical dashed line shows the overdensity at the virial radius $r_{200}$ of isothermal halos, $\Delta(r_{200})=200/3\approx67$, relevant if the gas within collapsed halos is to be excluded.
}
\label{clumping figure}
\end{figure}

\subsection{Clumping Factor}
\label{clumping factor}
As recombinations result from two-particle collisions, their average rate is proportional to the clumping factor $C\equiv \langle n^{2} \rangle/\langle n \rangle^{2}$ of the gas.
The clumping factor is therefore a key quantity in calculating the ionization history of the Universe.
We pause to discuss its numerical value.

Many early calculations of the evolution of the global ionized fraction during reionization \citep[e.g.,][]{1999ApJ...514..648M} assumed clumping factors derived from averaging over the entire gas content of high-resolution simulations \citep[e.g.,][]{1997ApJ...486..581G, 2003MNRAS.339..312S}, which were as large as $\sim30$ at the redshifts $\sim10$ relevant for hydrogen reionization.
Such clumping factors are however unrealistically high, as they include gas in collapsed halos which are associated with the galaxies that produce the ionizing photons.
As the escape fraction $f_{\rm esc}$ already accounts for the effects of clumpy gas within these sources, the latter should not be included in calculating the clumping factor used to calculate the recombinations in the diffuse IGM.

Let us identify the size of collapsed structures with their virial radius, with the latter approximated by the radius within which the mean density is 200 times the mean cosmic density \citep[e.g.,][]{2001PhR...349..125B}.
Then if halos are further approximated as isothermal spheres, $\rho(r)\propto r^{-2}$, their dimensionless overdensity $\Delta\equiv\rho/\langle \rho \rangle$ at the virial radius is given by $\Delta(r_{200})=200/3\approx67$.
A crude estimate of the clumping factor excluding gas within the virial radii of halos is then obtained by calculating $C\equiv \langle \Delta^{2} \rangle/\langle \Delta \rangle^{2}$ using the \cite{2000ApJ...530....1M} PDF truncated at $\Delta_{\rm max}=\Delta(r_{200})$.
Figure \ref{clumping figure} illustrates the result of this calculation at $z=2$,~4,~and~6 for varying $\Delta_{\rm max}$.
As expected, the clumping factor decreases with increasing redshift for fixed $\Delta_{\rm max}$.

For the redshifts $z>6$ relevant for hydrogen reionization, $C\lesssim4$ for $\Delta_{\rm max}=\Delta(r_{200})$.
If HeII reionization proceeds between at $z\sim3-4$, then slightly higher clumping factors $C\sim7$ are more appropriate.
Of course, these estimates are rough and we will therefore present our calculations for a range of reasonable clumping factors.

\subsection{Constraints from the Reionization of Hydrogen and Helium}
\label{reionization constraints}
The facts that hydrogen is reionized by $z=6$ and helium is fully ionized by $z\approx3$ constrain the production of ionizing photons at earlier times.
We investigate these constraints here.

The rate of change of the mass-weighted ionized fraction of species $i$ can be written as
\begin{equation}
\label{ionized fraction rate of change}
\frac{dx_{i}}{dt}
=
\frac{1}{n_{i}}
\int_{\nu_{i}}^{\infty}
\frac{d\nu}{h\nu}
\epsilon_{\nu}
- C R_{i}(T) n_{e} x_{i},
\end{equation}
where $R_{i}$ is the recombination coefficient from the ionized to the ground state.
Although many studies of reionization use the case-B recombination coefficient on the basis that recombinations directly to the ground state produce an ionizing photon and therefore have no net effect on the ionized fraction, most of the recombinations in a clumpy universe occur near dense regions (such as Lyman-limit systems), so that the reemitted photon is likely to be absorbed before it escapes into the diffuse IGM \citep[][]{2003ApJ...597...66M, 2007arXiv0711.1542F}.
We therefore use case-A recombination coefficients.

\begin{figure}[ht]
\begin{center}
\includegraphics[width=0.49\textwidth]{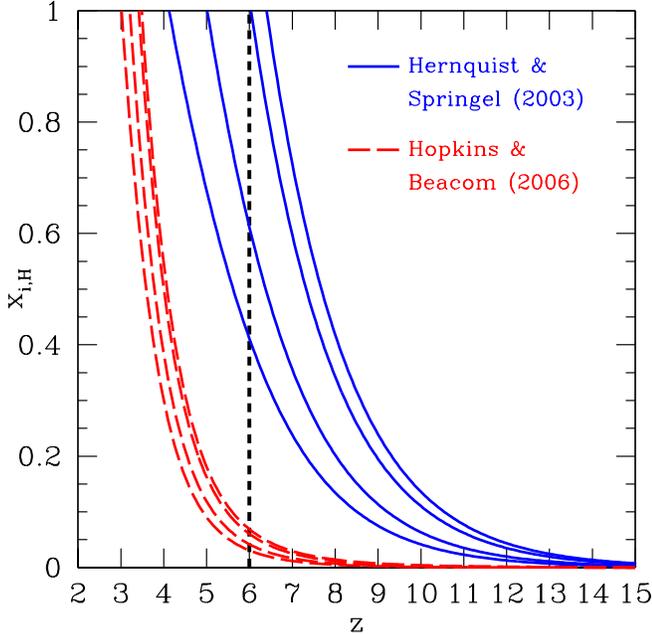}
\end{center}
\caption{
Evolution of the hydrogen ionized fraction.
Star-forming galaxies are assumed to be the dominant sources of ionizing photons, with emissivity at $z=3$ normalized to the value inferred from the hydrogen photoionization rate $\Gamma$ measured from the \Lya~forest.
The redshift evolution of the emissivity is assumed to follow the star formation rate density in the theoretical model of \cite{2003MNRAS.341.1253H} (blue solid) or the fit to observational data points compiled by \cite{2006ApJ...651..142H} (red dashed).
In each case, the clumping factor takes the values $C=0,~1,~5,$ and 10 from right to left.
The vertical line indicates redshift $z=6$, at which time the Universe is known to be reionized from observation of high-redshift quasars.
Even in the most optimistic case of no recombinations ($C=0$), the Universe fails to be reionized by this redshift by a large factor if the emissivity traces the star formation rate density fit of \cite{2006ApJ...651..142H}.
}
\label{h reion}
\end{figure}

\subsubsection{Hydrogen Reionization}
\label{hydrogen reionization}
We first consider the case of hydrogen reionization.
We assume that helium is singly ionized to HeII simultaneously ($x=y_{\rm II}$) with hydrogen, with both fully neutral at very high redshifts unaffected by reionization ($z\gtrsim50$).
%the initial redshift $z=50$.
We assume that stars are solely responsible for reionizing hydrogen and that the emissivity evolves with redshift following the SFR density estimates (eq. \ref{emissivity propto sfr}).
Beyond $z=3.9$, the highest redshift at which \cite{2001ApJ...557..519Z} measured the IGM temperature, we take the temperature to be constant at 22,000 K.
We normalize the value of the specific emissivity $\epsilon_{\nu}$ at the Lyman limit to be that inferred from $\Gamma$ measured from the \Lya~forest at $z=3$ using equation (\ref{epsilon LL from Gamma}), $\epsilon_{\nu_{\rm HI}}(z=3)=9.4\times10^{-50}$ comoving erg s$^{-1}$ Hz$^{-1}$ cm$^{-3}$.
The corresponding escape fraction will be estimated by comparing with the 1500~\AA\ luminosity density obtained by integrating over galaxies in \S \ref{direct comparison}.
For this conversion, we have assumed a spectral index $\alpha_{\rm HI}=-0.5$ shortward of the Lyman limit for a high-redshift ionizing background $J_{\nu}$ dominated by star-forming galaxies.
The theoretical starburst models of \cite{2001ApJ...556..121K} produce intrinsic spectra with approximately $F_{\nu}\propto \nu^{-1}$ between $\nu_{\rm HI}$ and $\nu_{\rm HeII}=4\nu_{\rm HI}$, covering the wavelength interval that contributes most significantly to $\Gamma$.
This spectrum is then hardened, $\alpha \rightarrow \alpha-1.5$, by IGM filtering.

The redshift evolution of the specific intensity is assumed to follow the SFR density.
Under the assumption that it does not evolve with redshift, this approach is free of assumptions in the escape fraction of galaxies.
We investigate a wide range of clumping factors by successively setting $C=0,~1,~5,~$and~10 (c.f. \S\ref{clumping factor}).

The results are shown in Figure \ref{h reion} for the theoretical model of \cite{2003MNRAS.341.1253H} and for the observational SFR history of \cite{2006ApJ...651..142H}.
Of particular note here is that the SFR density fit of \cite{2006ApJ...651..142H} (under the assumption we have made that it traces the ionizing emissivity) fails to reionize hydrogen by $z=6$ by a large factor even in the most optimistic case of no recombinations.
This further supports the conclusion that the \cite{2006ApJ...651..142H} SFR density estimates decline too steeply at high redshifts $z\gtrsim3$, which we had reached using only the measurement of the \Lya~opacity at $z\leq4.2$.

If the emissivity more closely follows the SFR density predicted by \cite{2003MNRAS.341.1253H}, then the Universe is reionized by $z=6$ provided that $C\sim1$.
This is lower than expected at $z=6$ according to our simple estimate in \S \ref{clumping factor}.
At higher redshifts more representative of the middle of reionization, however, the clumping factor is likely smaller.
In addition, if low-density gas is reionized first, then an effective clumping factor $C<1$ may result throughout most of reionization, because all the dense gas would remain locked up in self-shielded systems and should not be counted in the $\langle \Delta^{2} \rangle$ term \citep[][]{2000ApJ...530....1M}.
Even the \cite{2003MNRAS.341.1253H} model may be in tension with estimates of the redshift of reionization from the optical depth to Thomson scattering from the \emph{WMAP} 5-year data, which yield a sudden reionization redshift $z_{\rm reion}=11.0\pm1.4$ \citep[][]{2008arXiv0803.0586D}.
This tension could be eased if baryonic mass were more efficiently converted into ionizing photons at very high redshifts, for example as a result of a top-heavy IMF.
Alternatively, the gas consumption timescale could be shorter than the one fiducially assumed ($t^{\star}_{0}=2.1$ Gyr), which would push the peak of star formation to a higher redshift.

The conclusions of the present section are in general agreement with the similar analysis of \cite{2003ApJ...597...66M}, who found that the ionizing comoving emissivity cannot decline from $z=4$ up to $z\sim9$ by more than a factor 1.5 if reionization is to be complete at $z=6$.
Our results also support the picture of a ``photon-starved'' reionization similarly advocated by \cite{2007MNRAS.382..325B} based the $z=5-6$ \Lya~forest photoionization rate.

\begin{figure}[ht]
\begin{center}
\includegraphics[width=0.49\textwidth]{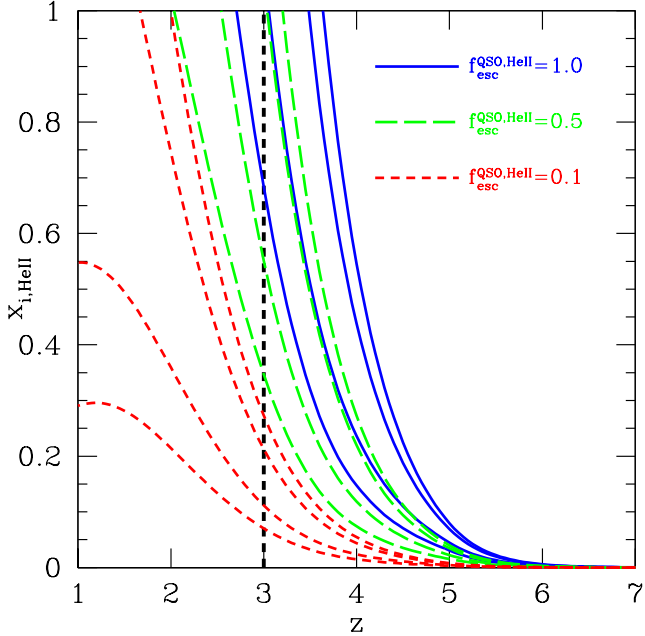}
\end{center}
\caption{
Evolution of the fraction of fully ionized helium.
All the helium is assumed to be initially in the form of HeII and to be doubly ionized by the action of quasars.
The quasar ionizing emissivity is calculated based on the $B$-band realization of the \cite{2007ApJ...654..731H} quasar bolometric luminosity function.
The escape fraction of HeII ionizing photons that escape their quasar source, $f_{\rm esc}^{\rm QSO,HeII}$, takes the values 1.0 (blue solid), 0.5 (green long dashed), and 0.1 (red short dashed). 
In each case, the clumping factor takes the values $C=0,~1,~5,$ and 10 from right to left.
The vertical line indicates redshift $z=3$, approximately the redshift at which observations of the HeII \Lya~forest provide evidence that HeII reionization is complete.
Escape fractions of HeII ionizing photons from quasars significantly less than unity are excluded in this scenario.
}
\label{heii reion}
\end{figure}

\begin{figure*}[ht]
\begin{center}
\includegraphics[width=1.0\textwidth]{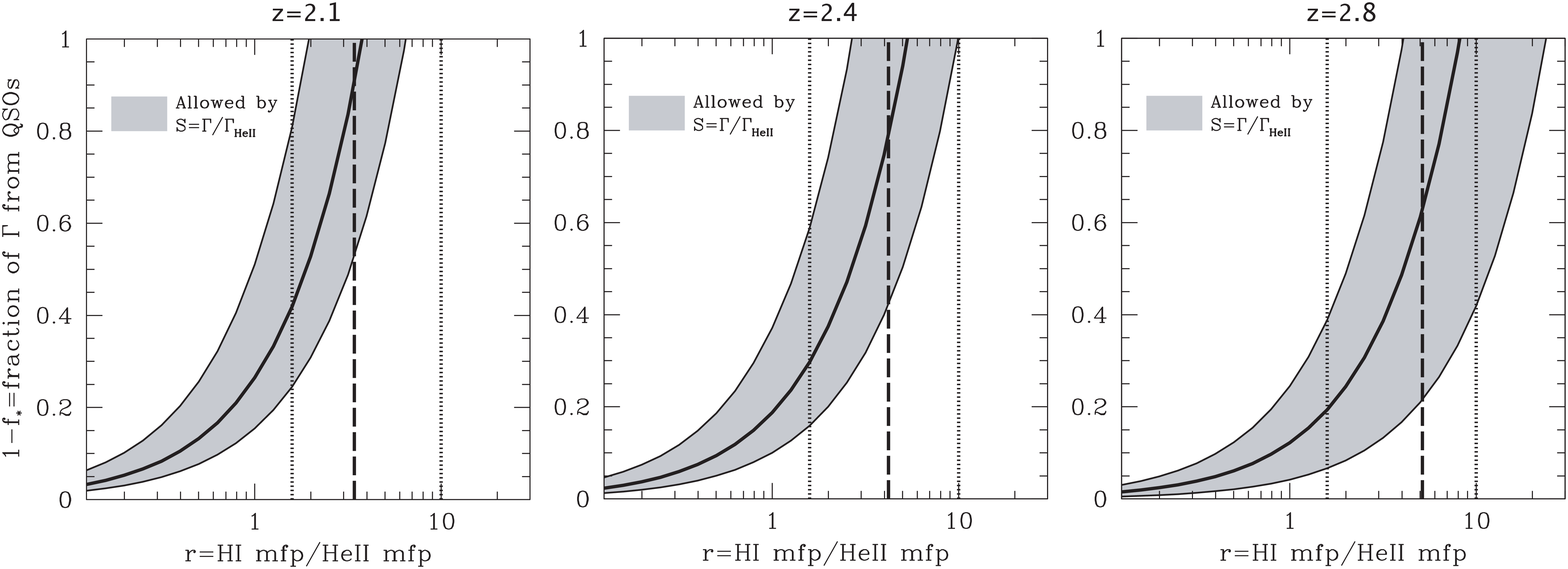}
\end{center}
\caption{
Fraction of the hydrogen total photoionization rate $\Gamma$ contributed by quasars ($1-f_{\star}$) as a function of the ratio of the mean free path of hydrogen ionizing photons to the mean free path of HeII ionizing photons.
The gray bands indicate the parameter space allowed by the measurement of the softness parameter of the UV background $S\equiv\Gamma/\Gamma_{\rm HeII}$ of \cite{2006MNRAS.366.1378B} at $z=2.1,~2.4,$ and 2.8.
The vertical long dashed lines show the ratio of the mean free paths estimated from the observed median column density ratio $\eta=N_{\rm HeII}/N_{\rm HI}$ at each redshift.
The vertical dotted lines indicate the range of plausible values for the ratio, defined as corresponding to $\eta=10$ and $\eta=400$, outside of which range measured values are probably not reliable \citep[][]{2004astro.ph.10588R, 2004ApJ...600..570S}.
}
\label{qsofrac all}
\end{figure*}

\subsubsection{Helium Reionization}
\label{helium reionization}
We also consider the reionization of HeII to HeIII, which we assume to proceed through the action of quasars only, requiring hard photons of energy $>54.4$ eV.
In this case, we can rewrite the emissivity integral as
\begin{equation}
\label{heii ionizing photons}
\int_{\nu_{\rm HeII}}^{\infty}
\frac{d\nu}{h\nu}
\epsilon_{\nu}
=
f_{\rm esc}^{\rm QSO,HeII}
\int_{0}^{\infty}
dL_{B} \dot{N}_{i, {\rm HeII}}
\frac{d\Phi}{dL_{B}},
\end{equation}
where
\begin{equation}
\dot{N}_{i, {\rm HeII}}=1.95\times10^{55}~\textrm{s}^{-1}
\left(
\frac{L_{B}}{10^{12}~L_{\odot}}
\right)
\end{equation}
is the number of HeII ionizing photons emitted per unit time by a quasar of $B$-band luminosity $L_{B}$ and $d\Phi/dL_{B}$ is again the $B$-band quasar luminosity function from \cite{2007ApJ...654..731H}.
We again investigate the wide range of clumping factors $C=0,~1,~5,$ and $10$.
We examine escape fractions of HeII ionizing photons from quasars $f^{\rm QSO,HeII}_{\rm esc}=0.1$ and $0.5$ in addition to unity.
The results are shown in Figure \ref{heii reion}.
We find that quasars are able to full ionize helium by $z\approx3$, as required by observations of the HeII forest, for escape fractions of unity.
Escape fractions $f^{\rm QSO,HeII}_{\rm esc}\lesssim0.5$ may however not be sufficient to fully reionize helium by this redshift unless the gas is unrealistically smoothly distributed ($C\approx1$; but see comment in \S \ref{hydrogen reionization} regarding the possibility of an effective clumping factor $C<1$ during reionization).\footnote{If the emergent quasar emission is beamed, then the true fraction of ionizing photons escaping into the IGM may be smaller. 
Since quasars beamed away from us would not be accounted for in the B-band luminosity function, this is however the relevant effective escape fraction to use in equation (\ref{heii ionizing photons}).}

An interesting consistency check on the contribution of quasars to the ionizing background is provided by measurements of the HI to HeII column density ratio of absorbers, $\eta \equiv N_{\rm HeII}/N_{\rm HI}$.
Let us assume that the total hydrogen photoionization rate $\Gamma$ is a sum of a stellar contribution and a quasar contribution, with a fraction $f_{\star}$ coming from stars:
\begin{equation}
\Gamma = \Gamma^{\star} + \Gamma^{\rm QSO};~\Gamma^{\star}=f_{\star}\Gamma~\&~\Gamma^{\rm QSO} = (1-f_{\star})\Gamma.
\end{equation}
Further assume that only quasars contribute significantly to the HeII ionizing budget, so that $\Gamma_{\rm HeII}=\Gamma^{\rm QSO}_{\rm HeII}$.
Again defining the helium terms in analogy to hydrogen, equation (\ref{epsilon LL from Gamma}) implies that
\begin{align}
\frac{\Gamma^{\rm QSO}}{\Gamma_{\rm HeII}} =&
\left( \frac{\epsilon_{\nu_{\rm HI}}^{\rm QSO}}{\epsilon_{\nu_{\rm HeII}}^{\rm QSO}} \right)
\left( \frac{\sigma_{\rm HI}}{\sigma_{\rm HeII}} \right)
\left( \frac{\alpha_{\rm HeII}^{\rm QSO}+3}{\alpha_{\rm HI}^{\rm QSO}+3} \right) \notag \\
& \times \left( \frac{\Delta l(\nu_{\rm HI})}{\Delta l(\nu_{\rm HeII})} \right),
\end{align}
where the superscripts indicate that only the quasar component is considered.
Defining the ratio of the mean free paths $r\equiv \Delta l(\nu_{\rm HI})/\Delta l(\nu_{\rm HeII})$ and for an unhardened emissivity $\epsilon^{\rm QSO}_{\nu}\propto \nu^{-\alpha^{\rm QSO}_{\rm unhard}}$, this reduces to
\begin{equation}
\frac{\Gamma^{\rm QSO}}{\Gamma_{\rm HeII}} = 
4^{\alpha^{\rm QSO}_{\rm unhard}+1} r,
\end{equation}
where we have used $\sigma_{\rm HI}=4\sigma_{\rm HeII}$ and $\nu_{\rm HeII}=4\nu_{\rm HI}$, and assumed that the UV background is uniformly hardened by IGM filtering, so that $\alpha_{\rm HeII}^{\rm QSO}=\alpha_{\rm HI}^{\rm QSO}$.
Combining the above expressions, we can solve for the fraction of the total $\Gamma$ contributed by quasars in term of the background softness parameter $S\equiv \Gamma/\Gamma_{\rm HeII}$ (where the HeII photoionization rate $\Gamma_{\rm HeII}$ is defined in analogy to the hydrogen rate $\Gamma$, but integrating only above the HeII photoelectric edge at 54.4 eV) and the ratio of the mean free paths:
\begin{equation}
1 - f_{\star} = \frac{4^{\alpha^{\rm QSO}_{\rm unhard}}r}{S},
\end{equation}\\ \\
As both $r$ and $S$ depend on $\eta$, measurements of the latter constrain $1 - f_{\star}$.
Using hydrodynamical simulations, \cite{2006MNRAS.366.1378B} infer the softness parameter $S$ from $\eta$ obtained by fitting spectra of the HI and HeII \Lya~forests of the quasar HE 2347$-$4342 \citep[][]{2004ApJ...605..631Z, 2004ApJ...600..570S}.
They find a softness parameter increasing from about $S=140$ at $z=2.1$ to about $S=300$ at $z=2.8$, with large uncertainties.
In Figure \ref{qsofrac all}, we show the contribution of quasars to the hydrogen photoionization rate $1-f_{\star}$ allowed by their measurement as a function of the ratio of the mean free paths $r$ at $z=2.1,~2.4$,~and~2.8.

What is a reasonable value for the ratio of the mean free paths?
Opacity owing to hydrogen only constrains the mean free path at the HeII photoelectric edge to be less than 8 times the value at the hydrogen Lyman edge (eq. \ref{mean free path equation}).
Helium opacity is however the limiting factor for the the mean free path of HeII ionizing photons.
A rough estimate of the HeII mean free path can be obtained by assuming a constant ratio of column densities $\eta=N_{\rm HeII}/N_{\rm HI}$ and making a Jacobian transformation to relate the column density of HeII to that of HI:
\begin{equation}
\frac{\partial^{2}N(N_{\rm HeII})}{\partial N_{\rm HeII}\partial z}
=\frac{1}{r}
\frac{\partial^{2}N(N_{\rm HI}=N_{\rm HeII}/r)}{\partial N_{\rm HI}\partial z}.
\end{equation}
Equation (\ref{mean free path equation}) may then be applied to express the HeII mean free path in terms of its column density distribution by simply making the replacement HI$\rightarrow$HeII.\footnote{This approach neglects the contribution of hydrogen opacity to the HeII mean free path. The latter is however negligible comparable to the HeII opacity for the measured column density ratios $\eta$.}
Taking the ratio to the mean free path at the Lyman edge owing to hydrogen opacity, we find
\begin{equation}
r=\left(\frac{4}{\eta}\right)^{1-\beta}.
\end{equation}

For the median column density ratios $\eta=(47,~68,~106)$ at $z=(2.1,~2.4,~2.8)$ \citep[][]{2004ApJ...605..631Z, 2006MNRAS.366.1378B} and approximating $\beta=1.5$, we find $r=(3.4,~4.2,~5.2)$.
These ratios are plotted as vertical dashed lines in Figure \ref{qsofrac all} and suggest that quasars contribute a large (perhaps most) of the hydrogen ionizing background at $z=2.1$, near the peak of the quasar luminosity function, with their contribution fractionally decreasing with increasing redshift.

We need to point out some uncertainties in this simple calculation.
There are three main sources.
First, in assuming a constant $\eta$ equal to the observed median, we have neglected the large fluctuations that are seen.
A rough estimate of the uncertainty in the ratio of the mean free paths induced by the large fluctuations is obtained by considering the extreme ratios $\eta=10$ and $\eta=400$, outside of which range the measured values are probably not reliable owing to difficulties introduced by background subtraction, the low signal-to-noise ratio of the spectra, line saturation, and blending of higher order Lyman series lines and metals \citep[][]{2004astro.ph.10588R, 2004ApJ...600..570S}.
These cases are indicated by the vertical dotted lines in Figure \ref{qsofrac all}.
Consistently, we have also used the softness parameter inferred by \cite{2006MNRAS.366.1378B} for a spatially uniform UV background, which will be in error if the column density fluctuations arise from a fluctuating radiation field.

Second, the vast majority of the absorbers for which the column density ratio $\eta$ has been measured are optically thin to both HI and HeII ionizing photons, whereas the systems that are optically thick to HeII ionizing photons are likely to contribute significantly to limiting their mean free path.
As the expected column density ratio $\eta$ varies non-monotonically with increasing column density ratio $N_{\rm HI}$ near the optically thick transition for reasonable background spectra \citep[e.g.,][]{1996ApJ...461...20H}, it is unclear in which direction our optically thin calculation for the mean free path ratio may be biased.

Finally, we have neglected a possible anti-correlation between $\eta$ and $N_{\rm HI}$ \citep[][]{2004astro.ph.10588R, 2004ApJ...600..570S}.
This anti-correlation would tend to make our calculation of the HeII ionizing mean free path an overestimate, so that $r$ would be underestimated.
In light of these uncertainties, the exact numerical values in Figure \ref{qsofrac all} should be interpreted with caution.
Nevertheless, the extreme cases indicated by the dotted lines and the steep increase of the quasar contribution $1-f_{\star}$ for $r>1$ convincingly argue that at their $z\sim2$ peak, quasars contribute a non-negligible fraction ($\gtrsim20$\%) of the hydrogen ionizing background, and possibly even dominate it.

A caveat to the above arguments is that it is not inconceivable that galaxies contribute a non-negligible fraction of HeII ionizing photons \citep[][]{2007arXiv0711.1542F}.
For example, the composite $z\sim3$ LBG spectrum of \cite{2003ApJ...588...65S} shows a significant HeII $\lambda$1640 recombination line but no evidence for AGN contamination.
In some scenarios, this Balmer-$\alpha$ line may be accompanied by HeII ionizing photons.
Associating the production of HeII ionizing photons with star formation, \cite{2007arXiv0711.1542F} show that a large escape fraction $\gtrsim50$\% for such photons is however required for LBGs to contribute comparably to quasars to the intergalactic HeII ionizing flux at $z\sim3$.
Although we cannot rule this out, the speculative nature of stars as a source of HeII ionizing photons and the large escape fraction that it would require motivate us to at present take the conservative view that this scenario is unlikely.
Moreover, a significant contribution to UV emission at the HeII photoelectric edge from hot gas in galaxies and galaxy groups \citep[][]{2004MNRAS.348..964M} would decrease the expected fluctuations in the HI to HeII column density ratio, which appears difficult to reconcile with the large increase in the HeII opacity fluctuations toward $z\gtrsim 3$ \citep[][]{2006MNRAS.366.1378B}.

\begin{center}
\begin{deluxetable*}{lcccccccl}
\tablewidth{0pc}
\tablecaption{Galaxy UV Luminosity Functions\label{UV luminosity functions table}}
\tabletypesize{\footnotesize}
\tablehead{\colhead{$z$} & \colhead{$M^{\star}$} & \colhead{$\phi^{\star}$} & \colhead{$\alpha$} & \colhead{Survey} & \colhead{Area} & \colhead{Reference} \\
\colhead{} & \colhead{} & \colhead{10$^{-3}$ com Mpc$^{-3}$} & \colhead{} & \colhead{} & \colhead{} & \colhead{}
}
\startdata

2.3\tablenotemark{a} & $-20.97\pm0.23$ & $1.74\pm0.63$ & -$1.84\pm0.11$ & Keck LBG & 4 deg$^{2}$ & \cite{2008ApJS..175...48R} \\
3.05\tablenotemark{a,b} & $-20.84\pm0.12$ & $1.66\pm0.63$ & -$1.57\pm0.22$ &  &  &  \\
4.0\tablenotemark{c} & $-20.84$ & $1.64\pm0.15$ & -$1.57$ &  &  & \cite{1999ApJ...519....1S} \\
\hline

1.7 & $-19.80\pm0.29$ & $17.0\pm4.3$ & $-0.81\pm0.18$ & Keck DF & 169 arcmin$^{2}$ & \cite{2006ApJ...642..653S} \\
2.2 & $-20.60\pm0.41$ & $3.02\pm1.79$ & $-1.20\pm0.23$ &  &  &  \\
3.0 & $-20.90\pm0.18$ & $1.70\pm0.42$ & $-1.43\pm0.13$ &  &  &  \\
4.0 & $-21.00\pm0.43$ & $0.85\pm0.65$ & $-1.26\pm0.38$ &  &  &  \\
\hline

3.8 & $-20.98\pm0.10$ & $1.3\pm0.2$ & $-1.73\pm0.05$ & Hubble UDF & 44 arcmin$^{2}$\tablenotemark{d} & \cite{2007ApJ...670..928B} \\
5.0 & $-20.64\pm0.13$ & $1.0\pm0.3$ & $-1.66\pm0.09$ & and other deep &  &  \\
5.9 & $-20.24\pm0.19$ & $1.4\pm0.5$ & $-1.74\pm0.16$ &  HST fields &  &  \\
\hline

4.0 & $-21.14\pm0.15$ & $1.46\pm0.38$ & $-1.82\pm0.09$ & Subaru DF & 875 arcmin$^{2}$ & \cite{2006ApJ...653..988Y} \\
4.7 & $-20.72\pm0.15$ & $1.23\pm0.36$ & $-1.82$\tablenotemark{e} &  &

\enddata
\tablecomments{Where the error is not explicitly given, the parameter was held fixed in the fit.}
\tablenotetext{a}{Central redshift of the sample.}
\tablenotetext{b}{Includes space-based Hubble Deep Field data.}
\tablenotetext{c}{Fit to the \cite{1999ApJ...519....1S} LBG data converted to $\Lambda$CDM cosmology. The $M^{\star}$ and $\alpha$ parameters were fixed to the $z\sim3$ determination of the luminosity function of \cite{2008ApJS..175...48R} (Naveen Reddy 2008, private communication).}
\tablenotetext{d}{Excluding the shallower GOODS fields also included in their analysis, but including the ``Parallel'' deep fields.}
\end{deluxetable*}
\end{center}

\subsection{Direct Comparison with Galaxy UV Luminosity Functions and the Escape Fraction of Ionizing Photons}
\label{direct comparison}
The arguments of this section have thus far fundamentally relied on the \emph{shape} of the redshift evolution of the quasar emissivity and of the total photoionization rate.
This is the reason we used analytic fits (which can be straightforwardly extrapolated to higher redshifts) to the star formation history, a quantity which at high redshifts is \emph{derived} from the observed UV luminosity functions.
As the UV luminosity functions themselves are the relevant quantities that involve the least unnecessary assumptions, it is of interest to compare them directly with the UV emissivity implied by our \Lya~forest measurement by equation (\ref{epsilon UV from epsilon LL}).

To estimate the contribution of star-forming galaxies to the UV background, we consider recent determinations of the galaxy UV luminosity function from Lyman break (LBG) surveys at $z\gtrsim2$ by \cite{2006ApJ...642..653S} (Keck Deep Fields), \cite{2007ApJ...670..928B} (Hubble Ultra Deep Field [HUDF] and other deep \emph{Hubble Space Telescope} [HST] deep fields), \cite{1999ApJ...519....1S} and \cite{2008ApJS..175...48R} (Keck LBG), and \cite{2006ApJ...653..988Y} (Subaru Deep Field).
These were selected to be the most up-to-date measurements in the fields covered.
The measured luminosity functions are summarized in Table \ref{UV luminosity functions table}, where in each case the parameters of the \cite{1976ApJ...203..297S} function are defined by
\begin{equation}
\phi(L)dL = \phi^{\star}
\left( \frac{L}{L^{\star}} \right)^{\alpha}
\exp{\left( 
-\frac{L}{L^{\star}}
\right)}
d\left(
\frac{L}{L^{\star}}
\right)
\end{equation}
and the break magnitude $M^{\star}$ is related to the characteristic luminosity $L^{\star}$ by the usual AB-magnitude conversion:
\begin{equation}
L^{\star} = 4 \pi (10~{\rm pc}/{\rm cm})^{2} 
10^{-0.4(M^{\star} + 48.60)}
{\rm~erg~s}^{-1}
\end{equation}
\citep[][]{1983ApJ...266..713O}.
Although the exact effective wavelength depends on the selection of the particular sample of interest, we will assume in our calculations that the UV luminosity functions are valid at 1500~\AA~and label the corresponding frequency by $\nu_{\rm UV}$.

For a UV luminosity function given in comoving units the corresponding comoving UV emissivity is simply obtained by taking the first moment of the luminosity function,
\begin{equation}
\epsilon_{\rm UV}^{\rm com} = 
\int_{L_{\rm min}}^{\infty}
dL L \phi(L) 
= L^{\star} \phi^{\star} \Gamma(\alpha+2, L_{\rm min}/L_{\star}),
\end{equation}
where only sources above the minimum luminosity $L_{\rm min}$ are accounted for and $\Gamma(\alpha+2, L_{\rm min}/L_{\star})$ is the upper incomplete gamma-function.

\begin{figure}[ht]
\begin{center}
\includegraphics[width=0.49\textwidth]{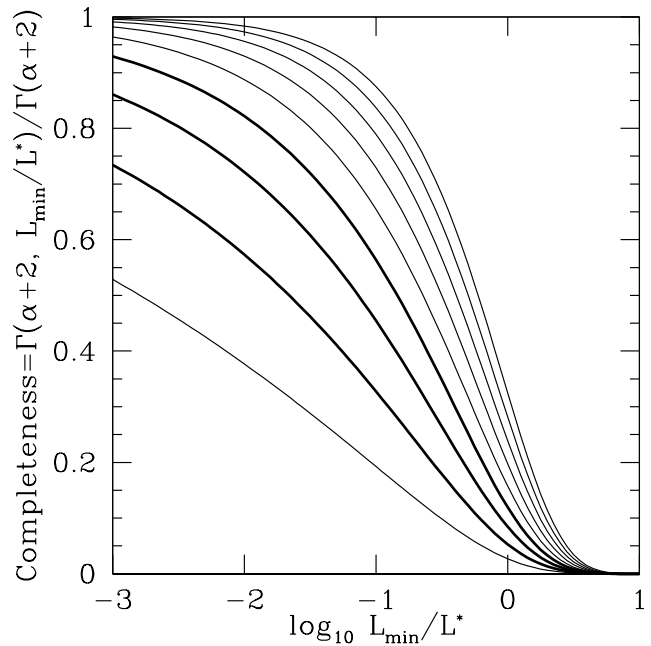}
\end{center}
\caption{
Completeness of the emissivity as a function of the fraction of the characteristic luminosity $L^{\star}$ down to which the luminosity function is integrated.
The different curves correspond to increasingly steep faint-end slopes from right to left: $\alpha=-1.1,~-1.2,~-1.3,~-1.4,~-1.5,~-1.6,~-1.7,~-1.8,$ and -1.9.
Slopes between -1.6 and -1.8 are typically measured (c.f. Table \ref{UV luminosity functions table}) and are indicated by thicker curves.
}
\label{gammainc}
\end{figure}

Figure \ref{gammainc} shows the completeness fraction $\Gamma(\alpha+2,~x)/\Gamma(\alpha+2)$ as a function of the faint-end slope $\alpha$ and the fraction of $L^{\star}$ down to which the luminosity function is integrated, $x\equiv L_{\rm min}/L_{\star}$. 
For steep faint-end slopes $\alpha\approx-1.8$ and surveys probing to $0.1L^{\star}$, only about one third of the UV emissivity is actually directly observed.
The extrapolation to the entire luminosity function ($L_{\rm min}=0$) thus generally involves a substantial factor.
It should therefore be borne in mind that a non-negligible error could be made if the faint-end slope is not well constrained, or changes below the present observational limits.

\begin{figure}[ht]
\begin{center}
\includegraphics[width=0.49\textwidth]{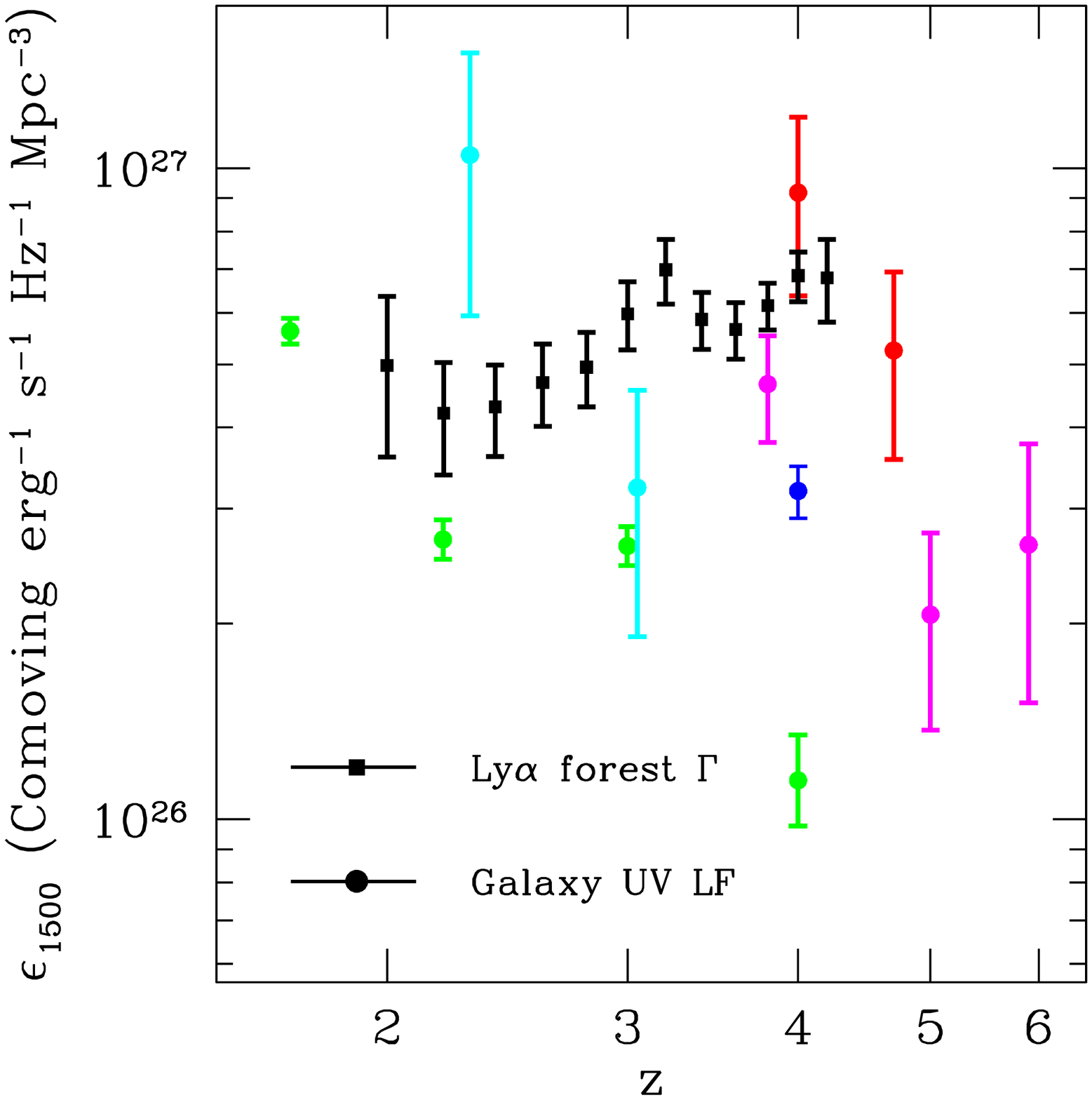}
\end{center}
\caption{Comoving UV specific emissivity at 1500~\AA~obtained by integrating galaxy UV luminosity functions.
The green points are from \cite{2006ApJ...642..653S} (Keck Deep Fields), the cyan points from \cite{2008ApJS..175...48R} (Keck LBG), the red points from \cite{2006ApJ...653..988Y} (Subaru Deep Field), and the magenta points from \cite{2007ApJ...670..928B} (Hubble Ultra Deep Field and other deep \emph{HST} fields).
The blue point corresponds to the luminosity function estimated from the \cite{1999ApJ...519....1S} $z\sim4$ LBGs, with the characteristic magnitude and faint-end slope set to the $z\sim3$ values of \cite{2008ApJS..175...48R} (Naveen Reddy 2008, private communication).
See the text for caveats about the error bars shown.
The black points show the emissivity implied by our \Lya~forest measurement, where the normalization has been set to match the luminosity function derived values.
The error bars on the emissivity implied by the \Lya~forest were artificially increased in the $\chi^{2}$ minimization to avoid giving unjustified weight to emissivities calculated from UV luminosity functions with unrealistically small error bars.
The error bars account only for the statistical uncertainty on our derived $\Gamma$ measurement only.
}
\label{epsilon UV figure}
\end{figure}

In Figure \ref{epsilon UV figure}, we show the UV emissivity from LBGs calculated by integrating the luminosity functions to $L_{\rm min}=0$ and the same quantity derived from our \Lya~forest measurement.
The error bars we quote on the emissivities calculated from the luminosity functions are propagated from those on the individual Schechter parameters; because the latter are generally correlated, these will overestimate the true errors on the luminosity densities.
Exceptions are the \cite{2006ApJ...642..653S} points, for which we take the total luminosity densities and errors reported by the authors.
As the photoionization rate $\Gamma$ is the quantity that is directly measured from the \Lya~forest, the normalization of the forest UV emissivity depends on the \emph{a priori} unknown ratio $(1/f_{\rm esc})(1500~{\rm \AA}/912~{\rm \AA})^{\alpha_{\rm UV}}(\alpha_{\rm HI}+3)$ (eqs. \ref{epsilon LL from Gamma} and \ref{epsilon UV from epsilon LL}).
We find the normalization minimizing the $\chi^{2}$ with respect to the UV emissivities calculated from the luminosity functions.
To do so, we consider only the data points overlapping with our measurement ($1.9\leq z \leq 4.3$) and compare them with the nearest $\Delta z = 0.2$ bin.
To avoid giving all the weight to a particular set of measurements, we artificially increase the error bars in this procedure.
The error bars on the emissivity implied by the \Lya~forest are artificially increased in the $\chi^{2}$ minimization to avoid giving unjustified weight to emissivities calculated from UV luminosity functions with unrealistically small error bars.

Solving for $f_{\rm esc}$ requires knowing $\alpha_{\rm UV}$ and $\alpha_{\rm HI}$.
We already discussed our choice of $\alpha_{\nu_{\rm HI}}=-0.5$ in \S \ref{hydrogen reionization}.
Rest-frame UV spectra of LBGs at $z\sim3$ \citep[e.g.,][]{2003ApJ...588...65S} are observed to be close to flat from 1500~\AA~ to \Lya~(1216~\AA).
Although the \Lya~forest suppresses the transmitted flux shortward of \Lya, there is little evidence for a change in the intrinsic galactic spectral shape down to the Lyman limit at 912~\AA.
We thus take $\alpha_{\rm UV}=0$.
Longward of the Lyman limit, the spectrum is not hardened by IGM filtering like ionizing photons.
Therefore, the spectral index of the UV background should be well-approximated by the spectral index of its sources.
Note that we are here assuming that $\Gamma$ is dominated by stellar emission over the entire redshift range probed by our \Lya~forest measurement.
If quasars contribute a large fraction at $z\sim2$, as suggested by the hardness of the ionizing background inferred from HeII column densities in \S \ref{helium reionization}, our estimate of the escape fraction will be biased high since a lower fraction of the ionizing photons would need to be accounted for by galaxies.
As the ionizing background can confidently be assumed to be dominated by stellar emission beyond $z\gtrsim3$ (\S \ref{comparison with forest}), our estimate should not be off by more than a factor of two because of this assumption.

The best-fit value for $f_{\rm esc}$ is 0.5\% for these values of $\alpha_{\rm UV}$ and $\alpha_{\rm HI}$.
Is this seemingly low value reasonable?

We need to pause and clarify our definition of $f_{\rm esc}$, as a number of variations are present in the literature.
Our definition as the number which satisfies equation (\ref{epsilon UV from epsilon LL}) measures the discontinuity of the emergent galactic spectrum at the Lyman limit.
Provided $\alpha_{\rm UV}\approx0$, as we have assumed, this definition is very close to the observational definition of \cite{2001ApJ...546..665S} as the ratio of the emergent specific flux from LBGs at 900~\AA~to that at 1500~\AA, $f[900]/f[1500]$.
This differs from the fraction of ionizing photons emitted by a galaxy that escape unabsorbed into the IGM owing to at least two effects.

First, a large fraction of UV photons at 1500~\AA~also do not escape the galaxy in which they are produced owing to absorption by dust.
This is the same effect that requires us to correct for dust absorption in estimating the rate of star formation from the observed UV luminosity functions (see \S \ref{dust corrections}).
For the ``common'' UV obscuration correction of \cite{2004ApJ...615..209H} at high redshifts, about 29\% of the intrinsic 1500~\AA~UV flux escapes its galaxy, but other estimates suggest that as little as 15-20\% may escape from $z\sim3$ LBGs \citep[][]{1998ApJ...508..539P, 2000ApJ...544..218A}.
This effect thus causes our $f_{\rm esc}$ to overestimate the true fraction of escaping photons by a factor $\sim5$.

Second, even in the absence of intergalactic absorption or absorption from the host galaxy, the intrinsic spectrum of the stellar population is likely to exhibit a break at the Lyman limit owing to the presence of hydrogen in the envelopes of the stars themselves.
If this is the case, then our $f_{\rm esc}$, which quantifies the magnitude of the Lyman-limit break, will underestimate the true fraction of ionizing photons that escape their galaxy by the same factor as the intrinsic stellar break.
The observational constraints on this intrinsic stellar break are essentially 
non-existent, as the effect is degenerate with galactic self-absorption and observations shortward of the Lyman limit are very challenging owing to the small amount of transmitted flux \citep[e.g.,][]{2001ApJ...546..665S, 2006ApJ...651..688S}.
Theoretical calculations based on stellar population synthesis however predict that the stellar Lyman-limit break can be a factor of $\sim3$ to $\sim5.5$ over a plausible range of ages and IMF \citep[][]{1999ApJS..123....3L, 2003MNRAS.344.1000B}.
This tends to counteract the overestimate caused by UV dust absorption, although the two effects are unlikely to perfectly cancel each other and a factor of $\sim2$ difference between our $f_{\rm esc}$ and the true faction of escaping ionizing photons is possible. 

From a composite spectrum of 29 $z\sim3$ LBGs, \cite{2001ApJ...546..665S} estimated a large $f_{\rm esc}\approx f[900]/f[1500]\approx20\%$.
However, as they note, the galaxies stacked in their composite spectrum were drawn from the bluest quartile of the LBG population and it is unclear whether the escape fraction of these objects is representative of all LBGs.
\cite{2006ApJ...651..688S} more recently obtained deep rest-frame spectroscopic Lyman continuum observations of 14 $z\sim3$ star-forming galaxies.
These authors detected escaping ionizing radiation from 2 individual objects in their sample, with the remaining 12 showing no evidence of Lyman continuum flux.
Averaging over their entire sample, they find an escape fraction $\sim4.5$ times lower than inferred by \cite{2001ApJ...546..665S}, $f_{\rm esc}\sim0.05$.
By measuring the incidence of absorbers optically thick at the Lyman limit in the spectra of 28 long-duration gamma-ray burst afterglows at $z\geq2$, \cite{2007ApJ...667L.125C} estimate the escape fraction at $0.02\pm0.02$, with a 95\% confidence level upper limit of 0.075.
Note that this latter escape fraction actually refers to the fraction of Lyman-limit photons that escape the host galaxy.

Another important point is that the escape fraction as measured from direct spectroscopic observations is inevitably representative of the bright-end of the galaxy population, since these observations are impossible on the faintest galaxies.
If the incidence of long-duration GRBs traces the SFR \citep[e.g.,][]{2008arXiv0801.3273F}, the GRB-DLA method will instead yield an escape fraction weighted by star formation, a large fraction of which occurs in fainter objects.
Even if this is not strictly the case, most GRBs appear to be hosted by relatively typical galaxies, the majority having relatively small mass and low luminosity \citep[][]{2007ApJ...657..367W, 2007ApJ...671..272C, 2008arXiv0802.0787L, 2008arXiv0803.2718S}.
Similarly, our method (solving for the value in eq. \ref{epsilon UV from epsilon LL} that brings the photoionization rate measured from the \Lya~forest in agreement with UV galaxy luminosity functions) will yield an escape fraction that is weighted by emissivity.
The numerical simulations of \cite{2007arXiv0707.0879G} suggest that the escaping ionizing photons originate from a small fraction of lines of sight essentially free of interstellar medium, as opposed to a nearly homogeneous semi-opaque medium.
If this is the case, then the escape fraction may well increase with the amount of galactic mechanical feedback, e.g. from supernova explosions \cite[][]{2002MNRAS.337.1299C, 2003ApJ...599...50F}.
Galaxies that are more luminous in the UV and therefore sustaining a higher SFR may thus have escape fractions higher than average.

Especially considering the discrepancy that could arise from the different definitions of the escape fraction, our small value is consistent (albeit on the low end) with the one derived by \cite{2007ApJ...667L.125C} from the GRB DLA distribution.
It is, however, significantly smaller than the values obtained by \cite{2001ApJ...546..665S} and \cite{2006ApJ...651..688S}.
By the remarks of the previous paragraph, the results could be reconciled if the escape fraction increases with galactic luminosity, with direct observations probing more luminous objects and our method and the GRB-DLA method probing fainter ones.
Although \cite{2001ApJ...546..665S} and \cite{2006ApJ...651..688S} predicted an ionizing background in reasonable agreement with values derived from the \Lya~forest for their high escape fractions,
these authors used values for the mean free path of ionizing photons at $z=3$ and an ionizing background spectral index $\alpha_{\rm HI}$ different from ours and only considered the emission from galaxies with UV luminosity above $0.1L^{\star}$, whereas we integrated the luminosity functions all the way to zero.
Each of these factors conspires to skew the calculation of the predicted $\Gamma$ in the same direction.
We have verified that our calculations are numerically consistent with those of \cite{2001ApJ...546..665S} and \cite{2006ApJ...651..688S} when the same mean free path, spectral indexes, and completeness are assumed.
The scaling of our derived escape fraction with these parameters is
\begin{align}
\label{fesc scaling}
f_{\rm esc}&=
0.5\%
\left(
\frac{\Delta l(\nu_{\rm HI},~z=3))}{85~{\rm proper~Mpc}}
\right)^{-1}
\left(
\frac{\alpha_{\rm HI}+3}{2.5}
\right)
\notag \\
&\times
\left(
\frac{912~{\rm \AA}}{1500~{\rm \AA}}
\right)^{-\alpha_{\rm UV}}
\left(
\frac{\Gamma(\alpha+2,~L_{\rm min}/L^{\star})}{\Gamma(\alpha+2)}
\right)^{-1} \notag \\
&\times\left(
\frac{\Gamma}{0.5\times10^{-12}~\rm s^{-1}}
\right).
\end{align}
Finally, it is interesting to note that this high-redshift estimate is in reasonable agreement even with estimates of the escape fraction from the $z=0$ Milky Way \citep[][]{1999ApJ...510L..33B, 2003ApJ...597..948P}.
A related question is whether this escape fraction is different for the earliest galaxies that are responsible for reionizing the Universe.

Within the large scatter, the redshift evolution of UV emissivity derived from the \Lya~forest is reasonably reproduced by the emission from LBGs only.
The only hint of a decline of the galaxy UV emissivity near $z=4$ comes from the highest-redshift point of \cite{2006ApJ...642..653S}.
This measurement is inconsistent with the higher points from the Subaru Deep Field \citep[][]{2006ApJ...653..988Y} and \cite{1999ApJ...519....1S}.
This may owe to cosmic variance in the relatively small Keck Deep Fields (KDF; 169 arcmin$^{2}$ vs. $\sim$850 arcmin$^{2}$ for Subaru and Steidel et al.), though in principle this should be mitigated by the three spatially independent fields of the KDF.
Quasars being clearly insufficient to solely account for the entire ionizing background implies that galaxies almost certainly dominate at $z\gtrsim3$.
Albeit with significant uncertainties, the fact that the total UV emissivity derived from the forest evolves similarly to the emissivity of LBGs suggests that the latter may in fact dominate the ionizing background at all redshifts $2\leq z \leq 4.2$ probed here.
The ratio of HeII to HI absorption however constrains the hardness of the UV background and limits the extent to which galaxies can dominate over quasars \cite[e.g.,][]{2001ApJ...549L.151H, 2006MNRAS.366.1378B}.

\section{THE MADAU DIAGRAM REVISITED}
\label{madau diagram}
Our results thus far have suggested tension between the state-of-the-art determinations of the instantaneous SFR density versus redshift (often known as the ``Madau diagram''; Lilly et al. 1996, Madau et al. 1996\nocite{1996ApJ...460L...1L, 1996MNRAS.283.1388M}), as represented by the fit of \cite{2006ApJ...651..142H}, and the cosmic UV luminosity density implied by the \Lya~forest.
In order to better understand the source of this discrepancy, we use recent UV luminosity functions to reconstruct the Madau diagram at $z\gtrsim2$ in a manner similar to previous work.
We then contrast this with what we obtain by deriving the diagram from the UV emissivity implied by our measurement of the \Lya~forest opacity.
The steps in this process are: 1) calculate the UV emissivity (either from from luminosity functions accounting for incompleteness, or from the \Lya~forest), 2) correct for dust extinction, and 3) convert from dust-corrected UV emissivity to SFR density.
These steps are particularly clearly summarized by \cite{2001MNRAS.320..504S} in their appendix.
We have already carried out the first step in the previous section (Figure \ref{epsilon UV figure}).

\subsection{Dust Corrections}
\label{dust corrections}
The quantity that is directly related to the SFR is the total number of UV photons produced by massive, short-lived stars.
A large fraction of these photons are destroyed by dust before they escape their host galaxy, so that the observed UV luminosity function must be corrected for dust obscuration before a SFR density can be inferred.
These corrections are very important, as they can be a factor of several. 

Unfortunately, we do not presently have an accurate handle on the dust corrections, which are likely to be at least luminosity-dependent \citep[][]{2001AJ....122..288H, 2003ApJ...599..971H, 2001ApJ...558...72S, 2003MNRAS.338..508P, 2003ApJ...597..269A, 2006ApJ...644..792R}.
Moreover, these should to some extent evolve with cosmic time as dust is produced and destroyed.
Here, we will make the simplest assumption that the UV attenuation factor is independent of both cosmic time and luminosity at the redshifts $z\gtrsim2$ we are concerned with.
Specifically, we will follow the ``common'' obscuration correction of \cite{2004ApJ...615..209H}, who assume the \cite{2000ApJ...533..682C} reddening curve resulting in an effective constant UV attenuation of a factor of 3.4.
For comparison, \cite{1999ApJ...519....1S} applied a correction factor of 4.7 at $z\sim3$ and $z\sim4$ based on the \cite{1997AJ....113..162C} reddening curve with typical $E(B-V)=0.15$.

Although this assumption is certain to be inexact, it has the advantage of being consistent with our analysis of \S \ref{comparison with forest}, where we investigated the cases of the ionizing emissivity tracing different SFR density evolutions.
It is also the same correction as applied by \cite{2006ApJ...651..142H} at high redshifts ($z\gtrsim3$) and therefore allows for a consistent comparison with the fit to the star formation history of these authors.
Moreover, the observational evidence for strong evolution of the high-redshift UV attenuation remains marginal and it is unclear whether a more sophisticated model is warranted at the present time.
For instance, \cite{2004ApJ...611..660O} find no evolution in dust extinction (based on E(B-V)) between LBGs at $z=3$ and 4.
The UV dust attenuation factor $\sim1.5$ derived by \cite{2007ApJ...670..928B} from the blue-slope $\beta-$values for $z\sim6$ $i-$dropouts \citep[][]{2005MNRAS.359.1184S, 2005ApJ...634..109Y, 2006ApJ...651...24Y} and the infrared excess-$\beta$ relationship \citep[][]{1999ApJ...521...64M} however suggests that the dust correction does eventually diminish with increase redshift.

In view of the clear uncertainties inherent in the dust corrections, we emphasize that our (and in fact all UV-derived) SFR density estimates should be interpreted with caution.
Our philosophy in this work is to explore the implications of the simplest assumptions which appear consistent with the existing evidence.

\subsection{From UV Luminosity to Star Formation Rate}
The basic conversion from specific UV luminosity in erg s$^{-1}$ Hz$^{-1}$ to SFR in $M_{\odot}$ yr$^{-1}$, which also directly converts from comoving specific UV emissivity to comoving SFR density, is that of \cite{1998ARA&A..36..189K}:
\begin{equation} 
\dot{\rho}_{\star} = 1.4\times10^{-28} \epsilon_{\rm UV}^{\rm com},
\end{equation}
where $\dot{\rho}_{\star}$ is in units of comoving M$_{\odot}$ yr$^{-1}$ Mpc$^{-3}$ provided $\epsilon_{\rm UV}^{\rm com}$ is expressed in erg s$^{-1}$ Hz$^{-1}$ Mpc$^{-3}$.
This conversion is appropriate for a \cite{1955ApJ...121..161S} IMF between 0.1 and 100 M$_{\odot}$ and a constant SFR of $\gtrsim100$ Myr is assumed.
In order to consistently compare with the \cite{2006ApJ...651..142H} SFR density fit for a modified Salpeter A IMF with a turnover below 1 M$_{\odot}$ \citep[][]{2003ApJ...593..258B}, we multiply the star formation rate by a further factor of 0.77.

\begin{figure}[ht]
\begin{center}
\includegraphics[width=0.49\textwidth]{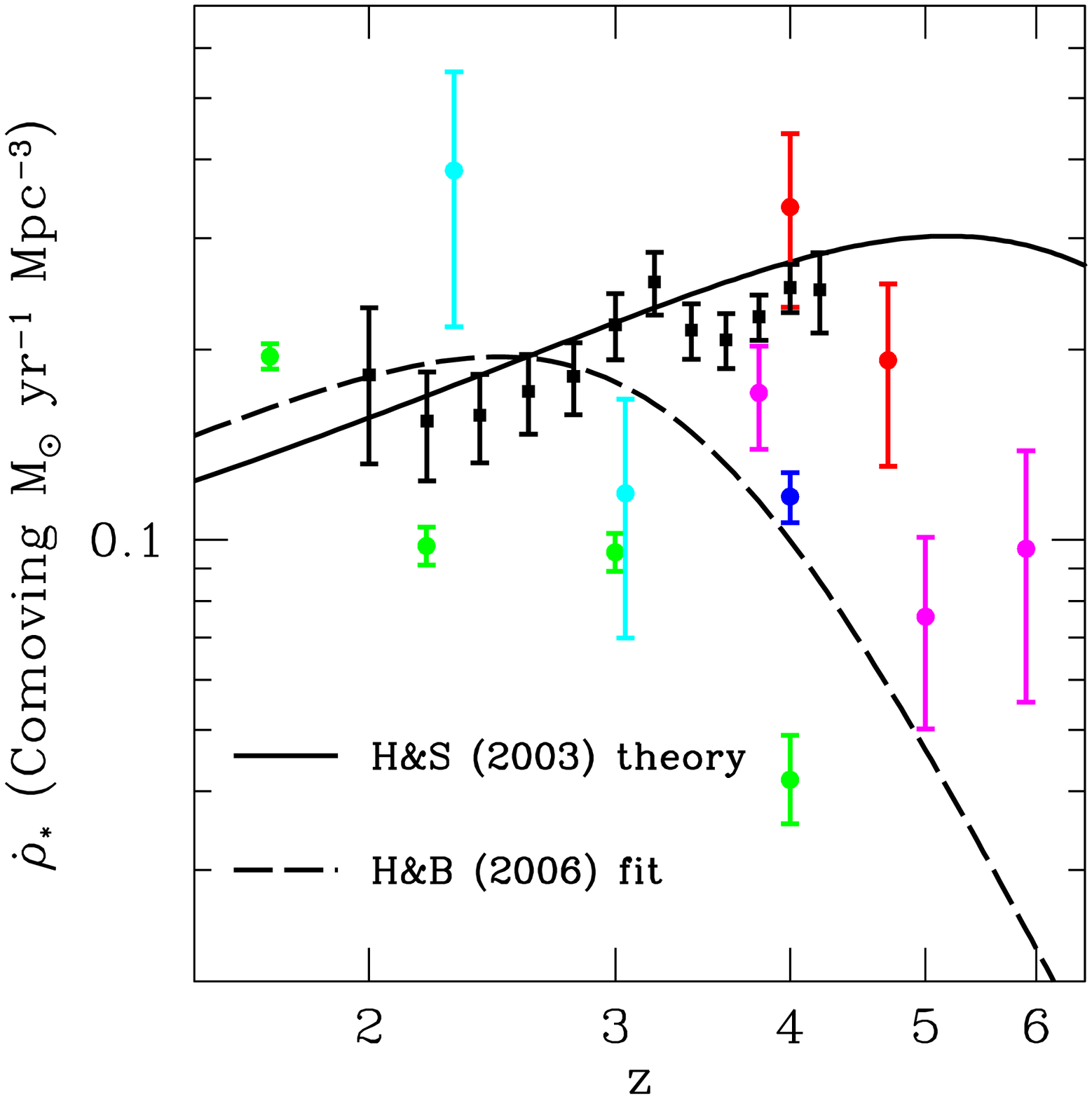}
\end{center}
\caption{
Comoving star formation rate density obtained with the \cite{1998ARA&A..36..189K} conversion from UV luminosity to SFR, modified for a Salpeter IMF with a turnover below 1 M$_{\odot}$ \citep[the modified Salpeter A IMF of][]{2003ApJ...593..258B}.
A constant dust obscuration correction factor of 3.4 has been applied at all redshifts following the high-redshift ``common'' correction of \cite{2004ApJ...615..209H}.
Same color scheme as in Figure \ref{epsilon UV figure}.
The dashed curve shows the best fit of \cite{2006ApJ...651..142H} to star formation history derived from their compilation of luminosity functions for a consistent IMF.
}
\label{sfr figure}
\end{figure}

\subsection{Star Formation Rate Results}
\label{sfr results}
In Figure \ref{sfr figure}, we show both the SFR density derived from galaxy UV luminosity functions and from the UV emissivity derived from our \Lya~forest opacity measurement, assuming the best-fit escape fraction of \S \ref{direct comparison}.
We find no compelling evidence for a decline in the comoving SFR density over the redshift range probed by our measurement, either from it or from the directly measured UV luminosity function, in contrast to the best fit of \cite{2006ApJ...651..142H}.
Although the star formation rate derived from the \Lya~forest is subject to a significant uncertainty on the mean free path of ionizing photons, our assumption for the redshift evolution of the latter is conservative and the discrepancy would be exacerbated if Lyman limit systems instead evolved more similarly to optically thin systems (\S \ref{mean free path section}). 

Inspection of Figure \ref{sfr figure} suggests that the present data are instead roughly consistent with a constant $\dot{\rho}_{\star}\sim0.2$ \mbox{M$_{\odot}$ yr$^{-1}$ Mpc$^{-3}$} at $2\lesssim z \lesssim 4.5$.
Since our analysis assumes a dust correction consistent with these authors at high redshifts, but is based on more recent data, it thus seems that (in combination with the arguments given in \S \ref{comparison with forest} and \ref{hydrogen reionization}) the SFR density peak suggested by their fit is an artifact of the scarce high-redshift data in their compilation, which may be affected by cosmic variance and is not uniformly complete.
For example, one of the $z\sim6$ points that drive the \cite{2006ApJ...651..142H} fit is the estimate of \cite{2004MNRAS.355..374B}, which is only complete to $0.1L^{\star}$ and is based on an extremely small HUDF 11 arcmin$^{2}$ exposure.
We instead consider the analysis of \cite{2007ApJ...670..928B}, which includes the HUDF data as a subset and yields a higher SFR density, and consistently integrate the luminosity functions down to zero luminosity.

The theoretical star formation history of \cite{2003MNRAS.341.1253H}, peaking between $z=5$ and $z=6$, appears to trace the observational points better up to $z\approx5$, though it may overestimate higher redshift points.
Recall, however, from \S \ref{hydrogen reionization} that the ionizing emissivity cannot decline much faster at high redshifts than predicted by the \cite{2003MNRAS.341.1253H} model in order for the Universe to be reionized by $z=6$.
If intrinsically faint galaxies are increasingly dominant at high redshift, then it is possible than even the current deepest surveys are missing a significant fraction of the star formation rate density toward $z\gtrsim5$.
We will return to this point in \S \ref{very faint galaxies}.

Note that, as in \S \ref{direct comparison}, we have again assumed that the hydrogen ionizing background is dominated by galaxies over the redshift range probed by our \Lya~forest measurement.
If quasars contribute significantly at $z\sim2$ (as suggested by HeII column densities; \S \ref{helium reionization}), the predicted star formation rate from the \Lya~forest in Figure \ref{sfr figure} should be lowered at these redshifts.
As stars must dominate beyond $z\gtrsim3$ (\S \ref{comparison with forest}), our remarks regarding star formation are unaffected at these redshifts.
In Figure \ref{sfr figure}, we have also renormalized the \cite{2003MNRAS.341.1253H} curve by a factor of  three to match the observational points.
Whereas the normalization of the observational points are subject to assumptions about the stellar IMF and dust corrections, the theoretical model is free of such assumptions and the two may therefore appear offset.
Moreover, a discrepancy of this magnitude also exists between the observationally measured stellar mass density versus redshift and the time integral of the star formation rate \cite[][]{2006ApJ...651..142H, 2008MNRAS.385..687W}, suggesting that at least one of them is incorrectly normalized.

It is immediately clear from the scatter in Figures \ref{epsilon UV figure} and \ref{sfr figure} that the total UV luminosity density extrapolated from the measured luminosity functions should be interpreted with caution.
In fact, the dispersion between different points at fixed redshift is generally larger than the calculated error bars, indicating that these are unlikely to be uniformly reliable, a situation which is particularly manifest at $z\sim4$.
There are several reasons why this may be the case.
First, in all cases, the estimated luminosity density relies on an extrapolation to fainter magnitudes than probed by individual surveys.
Second, cosmic variance arising from large-scale structure is difficult to accurately quantify and may not be properly accounted for in all the measurements reported.
Finally, some parameters are (perhaps inaccurately) held fixed in some fits.
For instance, the $z\sim4$ \cite{1999ApJ...519....1S} LBG point assumes that the characteristic magnitude and faint-end slope measured at $z\sim3$ by \cite{2008ApJS..175...48R}; only the density parameter $\phi^{\star}$ is fitted for in this case.
Until these discrepancies are resolved, it appears dangerous to attempt a detailed fit of the $z\gtrsim2$ SFR history.

\section{DISCUSSION}
\label{discussion}
\subsection{Reemission}
In addition to the Lyman-continuum photons directly emitted by quasars and galaxies, the IGM itself acts as a source through recombinations \citep[e.g.,][]{1996ApJ...461...20H}.
In the clumpy IGM, most of the ionizing photons are absorbed in Lyman-limit systems.
Most likely, the atom that is ionized will be hydrogen.
The free electron will then thermalize to a temperature $\sim10^{4}$ K with the local gas.\footnote{This gas need not have the same temperature as the diffuse IGM, but atomic cooling will prevent it from reaching much higher values in dense systems.}
Because the thermal kinetic energy of the electron is much smaller than the hydrogen ionization potential 13.6 eV$\approx1.6\times10^{5}$ K, even if the electron recombines directly to the ground state and releases an ionizing photon, the latter will have energy barely above the Lyman limit.
Background photons with initially high energy will thus penetrate deep into a Lyman-limit system, while the lower-energy remitted ionizing photons will remain trapped.
Thus, only background photons with energy just above the Lyman limit will be absorbed on the skin of the Lyman-limit system and have a significant probability that a reemitted Lyman-continuum photon escapes the Lyman-limit system.
\cite{2003ApJ...597...66M} estimates the increase of Lyman-limit emissivity owing to his effect to be $\lesssim$10\%.
The effect of recombinations to HeII is most important prior to HeII reionization, when even the diffuse IGM is optically thick to HeII ionizing photons, in which case \cite{2003ApJ...597...66M} estimates the effect at $\lesssim$4\%.
Therefore, reemission is likely to affect the hydrogen ionizing emissivity at only the $\lesssim10-15$\% level.
Since this is relatively small compared to the other uncertainties we are dealing with, we neglect this in our considerations.

\begin{figure}[ht]
\begin{center}
\includegraphics[width=0.49\textwidth]{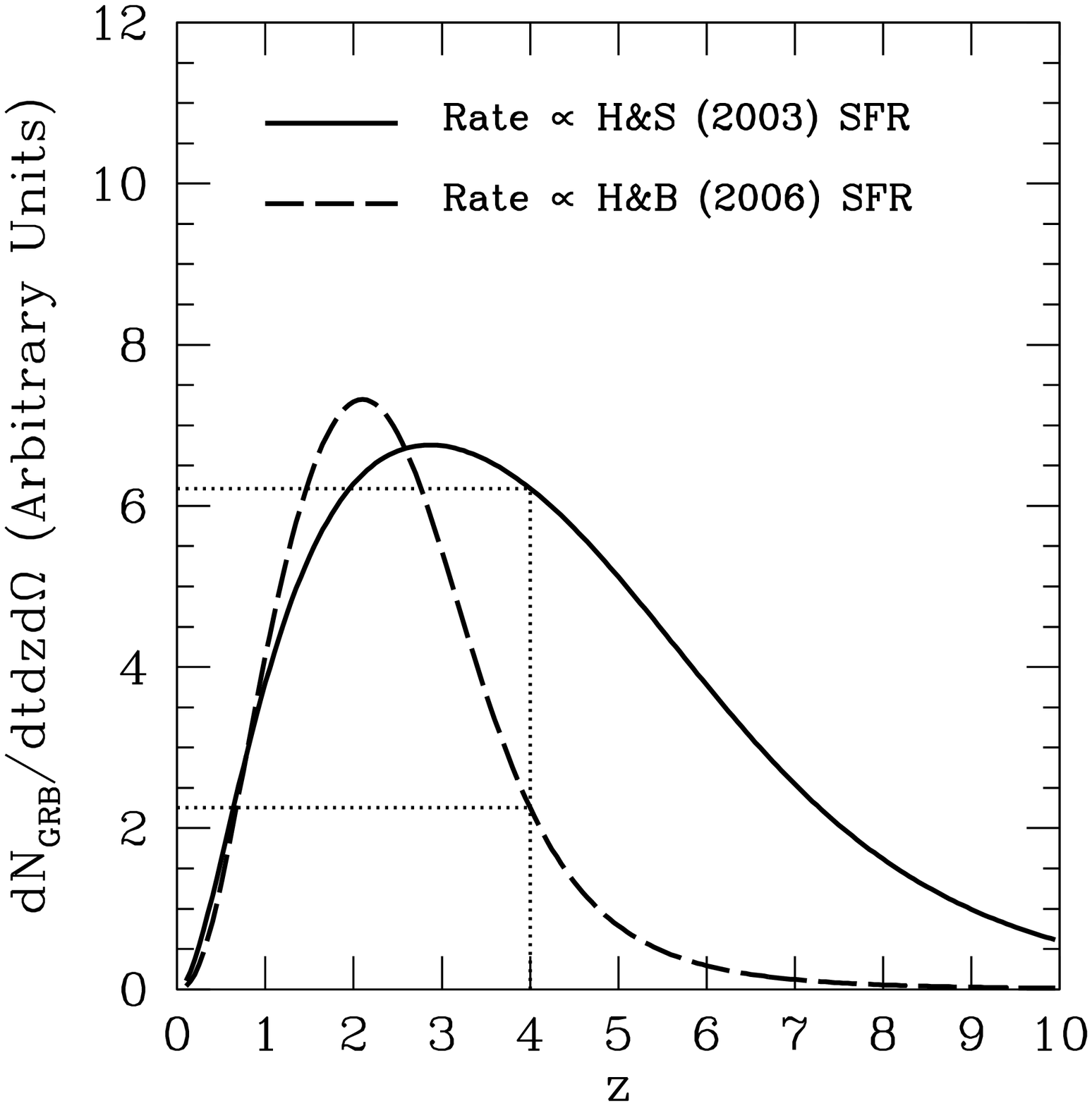}
\end{center}
\caption{
Predicted gamma-ray burst rate per unit redshift per steradian in arbitrary units, assuming that the GRB rate density follows the star formation rate density of \cite{2003MNRAS.341.1253H} (solid) and \cite{2006ApJ...651..142H} (long dashed) as shown in Figure \ref{sfr figure}.
The dotted lines indicate the rates at $z=4$, at which \cite{2008ApJ...673L.119K} find that the \cite{2006ApJ...651..142H} SFR underestimates the observed GRB rate by a factor of four.
The \cite{2003MNRAS.341.1253H} SFR model prediction is in better agreement with the observed GRB rate at this redshift.
}
\label{grb figure}
\end{figure}

\subsection{The Gamma-Ray Burst Rate}
\label{grb rate}
It is interesting to consider the gamma-ray burst rate under the assumption that it follows star formation.
In Figure \ref{grb figure}, we plot the predicted rate per unit redshift per unit steradian in arbitrary units\footnote{The absolute normalization of this rate is poorly constrained at present owing to the complicated selection function of the bursts.} under this assumption,
\begin{equation}
\label{grb rate observable eq}
\frac{dN_{\rm GRB}}{dt dz d\Omega} \propto \dot{\rho}_{\star}(z) 
\frac{c d^{2}(z)}{(1+z)H(z)},
\end{equation}
where $d(z)$ is the comoving distance to redshift $z$, for both the \cite{2003MNRAS.341.1253H} and \cite{2006ApJ...651..142H} SFRs shown in Figure \ref{sfr figure}.
Analyzing a sample of 36 luminous \emph{Swift} GRBs with redshift in the range $z=0-4$, \cite{2008ApJ...673L.119K} conclude that $\sim4$ more GRBs are observed at $z\approx4$ than expected if the GRB rate follows the \cite{2006ApJ...651..142H} SFR.
Interestingly, as the Figure shows, the \cite{2003MNRAS.341.1253H} model predicts more GRBs by a factor of $\sim3$ at this redshift, in better agreement with the observed rate.
\cite{2006MNRAS.372.1034D} and \cite{2007JCAP...07....3G} have similarly found that an enhancement of high-redshift GRBs is needed with respect to several existing estimates of the SFR.
Although it remains to be established whether the GRB rate actually traces star formation linearly, this provides further circumstantial evidence that the SFR does not decline strongly beyond $z\sim3$ as in the \cite{2006ApJ...651..142H} fit.
Along with the supporting arguments based on the IGM opacity that we have developed in this paper, this also indicates that it may be premature to draw conclusions about deviations of GRBs from star formation based on the present observational estimates.

\subsection{A Large Population of Very Faint Galaxies at High Redshifts?}
\label{very faint galaxies}
We found in \S \ref{hydrogen reionization} that a star formation history peaking early as in the \cite{2003MNRAS.341.1253H} model appears to be necessary to reionize the Universe by $z=6$.\footnote{Note that by normalizing to the ionizing emissivity implied by the \Lya~forest at $z=3$, our calculation implicitly assumes the small escape fraction $f_{\rm}\sim0.5\%$ also implied by the forest (\S \ref{direct comparison}).} 
Such a star formation history also accommodates the observed $z\approx4$ GRB rate better (\S \ref{grb rate}), in addition to providing an excellent fit to the intergalactic hydrogen photoionization rate now precisely measured at $z=2-4.2$ (\S \ref{comparison with forest}).
Yet, it appears to overestimate the SFR estimated from directly integrating the observed UV luminosity function at $z\gtrsim5$ \citep[][\S \ref{sfr results}]{2007ApJ...670..928B}. 
Can these results be reconciled?

An intriguing possibility is that the very high redshift Universe harbors a large number of extremely faint galaxies that are missed by even the deepest surveys to date \citep{2006ApJ...653..988Y, 2007ApJ...670..928B, 2008arXiv0803.0548B}.
As we have integrated the measured luminosity functions down to zero luminosity, this would require a significant steepening of the faint-end slope of the UV luminosity function below the present observational limits.
Faint galaxies may in fact already have been detected in abundance in the deep $z\sim6-10$ searches of \cite{2006A&A...456..861R} and \cite{2007ApJ...663...10S}, although the authenticity of the candidates remain challenging to confirm and their large number poses challenges to theoretical expectations \citep[][]{2007ApJ...668..627S}.
The recent detection of very faint Lyman emitters at $2.67 \leq z \leq 3.75$ by \cite{2007arXiv0711.1354R} that would have been missed in existing LBG surveys also reminds us that the very dim Universe may contain surprises. 

Ultimately, the viability of this possibility will be addressed by future, deeper surveys with instruments such as the \emph{James Webb Space Telescope}\footnote{http://www.jwst.nasa.gov/}, the \emph{Thirty Meter Telescope}\footnote{http://www.tmt.org/}, and the \emph{Giant Magellan Telescope}.\footnote{http://www.gmto.org/}

\subsection{The Sources of Reionization}
The decline of the quasar luminosity function and the increasing dominance of stellar emission to the high-redshift $z\gtrsim3$ ionizing background make a compelling case that the Universe was reionized by stars.
This gives credibility to analytical and numerical calculations of hydrogen reionization that make this assumption \citep[e.g.,][]{2003MNRAS.344..607S, 2004ApJ...613....1F, 2007MNRAS.377.1043M, 2007ApJ...654...12Z}.
This is encouraging news for upcoming observational probes of the epoch of reionization, such as redshifted 21-cm emission and high-redshift \Lya~emitters \citep[e.g.,][]{2004ApJ...608..622Z,2007MNRAS.381...75M,2008arXiv0802.1710M}, whose detailed interpretation will rely on our understanding of the morphology of a reionization and its origin.

\begin{figure}[ht]
\begin{center}
\includegraphics[width=0.49\textwidth]{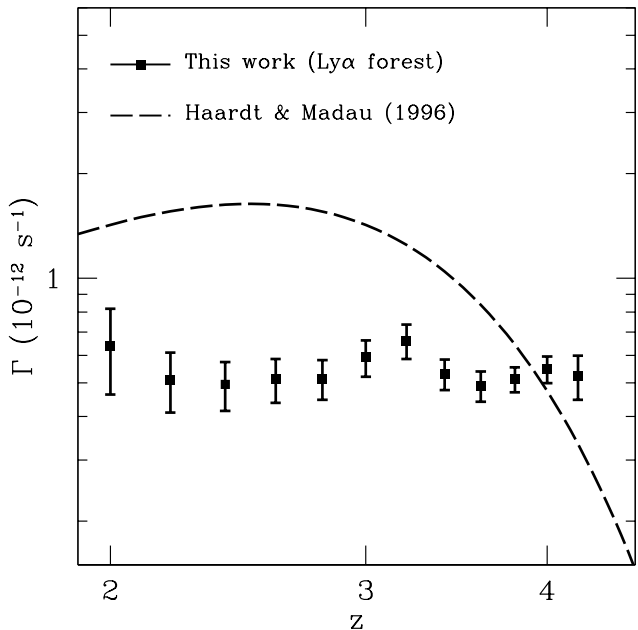}
\end{center}
\caption{
Hydrogen photoionization rate in the \cite{1996ApJ...461...20H} model (long dashed) compared to our empirical measurement from the \Lya~forest (squares).
}
\label{hmcomp fig}
\end{figure}

\subsection{Models of the Ionizing Background}
\label{haardt and madau}
Hydrodynamical cosmological simulations \citep[e.g.,][]{1996ApJ...457L..51H, 1996ApJS..105...19K, 1999ApJ...511..521D, 2003MNRAS.339..312S} require a model of the ionizing background in order to calculate the thermal evolution of the gas.
Most simulations presently adopt a model based on the pioneering work of \cite{1996ApJ...461...20H} or one of its variants.
In Figure \ref{hmcomp fig}, we compare the \cite{1996ApJ...461...20H} model with our empirical measurement from the \Lya~forest.
The model clearly deviates from our empirical measurement, highlighting the need for a new model of the UV background and its spectrum.
Although both are uncertain in normalization, the \cite{1996ApJ...461...20H} UV background is dominated by quasars, peaks at $z=2-3$, and subsequently declines steeply, in conflict with the measured intergalactic opacity, as we have argued at length.
In instances where only the photoionization rate $\Gamma$ is of interest, our empirical measurement will provide a valuable input to simulations.

\subsection{Comparison with Previous Work}
\label{previous work}
In many ways where there is overlap, our results echo and strengthen conclusions previously reached in a number of studies of the UV background \citep[e.g.,][]{1997ApJ...489....7R, 2001ApJ...549L..11M, 2005MNRAS.357.1178B}.
\cite{2005MNRAS.357.1178B}, in particular, inferred $\Gamma$ from the \Lya~opacity measurement of \cite{2003ApJ...596..768S} and also found its evolution to be consistent with being constant at $2\leq z\leq4$.
They found values of the photoionization rate $\Gamma\sim1\times10^{-12}$ s$^{-1}$ systematically higher than ours but with large error bars not inconsistent with them.
These authors performed a more detailed analysis of the sources of error in their measurement, including parameters of the cosmology and thermal state of the IGM, than we did here.
As the full thermal \emph{history} of IGM is possibly as important (see remarks in \S \ref{gas density pdf}) and remains an open and difficult question in itself, we have here inferred the photoionization rate in a relatively simplistic manner, focusing our efforts instead on the implications for its sources.
More work remains to be done to refine the absolute normalization of the ionizing background.
By comparing with the estimated quasar contribution, \cite{2004MNRAS.348L..43B} also found evidence for a stellar-dominated UV background at all redshifts.
Their argument was mostly based on the absolute value of $\Gamma$.
As its redshift evolution is constrained in much finer detail here, we have made use of its flat \emph{shape} in conjunction with the narrow peak of the quasar luminosity function to provide an argument that is more robust to normalization uncertainties.

Measurements based on the proximity effect \citep[][]{1987ApJ...319..709C, 1988ApJ...327..570B, 1991ApJ...367...19L, 1993ApJ...413L..63K, 1994ApJ...428..574W, 1995MNRAS.273.1016C, 1996ApJ...466...46G, 1996MNRAS.280..767S, 2000ApJS..130...67S} have tended to yield $\Gamma$ values higher by a factor of $\sim3$ (c.f. Figure \ref{gamma comparison}).
However, the overdense regions in which quasars reside are likely to bias these measurements high \cite[][]{1995ApJ...448...17L, 2008ApJ...673...39F}.
Moreover, many previous studies of the proximity effect used quasar systemic redshifts based on broad emission lines, which are increasingly recognized as biased measures as a result of inflows and outflows of material \citep[][]{1982ApJ...263...79G, 1992ApJS...79....1T, 2001AJ....122..549V, 2002AJ....124....1R, 2006ApJ...651...61H, 2007ApJ...655..735H}.
In contrast, the flux decrement method is both easier to model theoretically and prone to fewer observational challenges.
Moreover, it provides much better statistics, as the \Lya~forest typically covers a region away from the quasar much longer than its proximity region.
We therefore regard it as the more robust method of measuring $\Gamma$ from the forest.

A number of previous studies of the LBG UV luminosity function have also found little evidence for a decline of the SFR density beyond $z\sim3$ \citep[e.g.,][]{1999ApJ...519....1S, 2004ApJ...600L.103G, 2006ApJ...653..988Y, 2008ApJS..175...48R}.
This finding has in addition been corroborated by measures based on photometric redshifts \citep[e.g.,][]{2001ApJ...546..694T, 2003ApJ...596..748T}.
Recently, \cite{2008ApJ...675...49S} have also come to the conclusion that quasars can contribute at most about half of the total photoionization rate at $z\sim3$ from an optical/infrared selection of quasars in the Spitzer Wide-area Infrared Extragalactic ($SWIRE$) Legacy Survey, though their argument relies on the uncertain conversion from a luminosity function to an ionization rate (see \S \ref{comparison with forest}).

If only because the SFR density is expected to rise continuously on physical grounds, it must eventually decline toward high redshifts.
\cite{2007ApJ...670..928B} in fact find evidence for such a decline toward $z=6$ on the basis of evolving dust obscuration suggested by observed $\beta$-values at this redshift \citep[e.g.,][]{2005MNRAS.359.1184S}.
We simply contend here that neither the present \Lya~forest data nor the recent UV luminosity functions compiled here, especially when considered together with their mutual scatter, show convincing evidence for the often-assumed peak in SFR density near $z\sim2-3$.

\section{CONCLUSIONS}
\label{conclusions}
We have investigated the implications of the evolution of the intergalactic opacity for the sources responsible for its photoionization.
The principal constraint is our measurement of the \Lya~effective optical depth over the redshift range $2\leq z \leq4.2$ from a sample of 86 high-resolution quasar spectra \citep[][]{2008ApJ...681..831F}.
We also imposed the requirements that intergalactic HI must be reionized by $z=6$ and HeII by $z\approx3$, and used estimates of the hardness of the ionizing background from HI to HeII column density ratio measurements.
Our main conclusions are as follows:
\begin{enumerate}
\item The intergalactic hydrogen photoionization rate is remarkably constant over the redshift range $2 \leq z \leq 4.2$, with a value $\Gamma \approx (0.5\pm0.1)\times 10^{-12}$ s$^{-1}$, subject to remaining systematic uncertainties in normalization.
\item Because the quasar luminosity function is strongly peaked near $z\sim2$, the lack of redshift evolution implies that star-forming galaxies are likely to dominate the photoionizing background at $z\gtrsim3$, and possibly at all redshifts probed.
\item Although our arguments are robust to normalization systematics, fiducial assumptions regarding the mean free path of ionizing photons, the spectral energy distribution of quasars, and an escape fraction of ionizing photons from quasars of unity imply that quasars alone overproduce the total ionizing background near their peak. This puzzle suggests that these assumptions warrant more scrutiny and highlights the uncertainties in calculating the ionizing background by integrating luminosity functions. 
\item Only a small escape fraction of ionizing photons $f_{\rm esc}\sim0.5\%$ is needed for galaxies to solely account for the entire UV luminosity density implied by the \Lya~forest, with the scaling with uncertain parameters given in equation \ref{fesc scaling}.
This small fraction might be reconciled with higher measurements from the direct detection of Lyman-continuum photons from high-redshift galaxies \citep[][]{2001ApJ...546..665S, 2006ApJ...651..688S} if it increases with galaxy luminosity.
\item The state-of-the-art observational fit to the cosmic star formation history by \cite{2006ApJ...651..142H} (peaking near $z\sim2$ similarly to quasar activity) appears to underestimate the total photoionization rate by almost an order of magnitude at $z\approx4$ if the escape fraction and dust obscuration are constant with redshift for $z\gtrsim2$, and is also in tension with the most recent high-redshift determinations of the galaxy UV luminosity function.
\item Normalizing the ionizing emissivity predicted by the best-fit star formation history of \cite{2006ApJ...651..142H} to our \Lya~forest measurement at $z=3$, their star formation history fails to reionize the Universe by $z=6$.
\item A star formation history peaking at a higher redshift $z=5-6$, like the theoretical model of \cite{2003MNRAS.341.1253H}, fits the $2\leq z\leq 4.2$ \Lya~forest well, reionizes the Universe in time, and is in better agreement with the rate of GRBs observed by \emph{Swift} at $z\approx4$.
\item Quasars alone suffice to doubly ionize helium by $z\approx3$ provided that most of the HeII ionizing photons they produce escape into the IGM.
\item The ratio the HI to the HeII column densities in the post-HeII reionization epoch $2\lesssim z \lesssim3$ suggests, with a significant uncertainty arising from the large fluctuations observed, that quasars contribute a non-negligible fraction ($\gtrsim20$\%) and perhaps dominate the hydrogen ionizing background at their $z\sim2$ peak.
\end{enumerate}

Although the present uncertainties in the redshift evolution of UV dust corrections, the escape fraction of ionizing photons from galaxies, and possible evolution of the stellar initial mass function however prohibit definitive conclusions about the cosmic SFR to be reached at this time, we have illustrated how the IGM represents a powerful integral probe of the cosmic luminosity density free of the completeness issues that affect the direct detection of its sources.
In this sense, its study is very complementary to that of luminosity functions.

In this work, we have strived to make the simplest assumptions consistent with our present body of knowledge.
For instance, we have assumed that the escape fraction of ionizing photons from star-forming galaxies is constant with redshift and found evidence that the high-redshift ($z\gtrsim3$) SFR density is currently underestimated by direct galaxy counts.
Of course, many of our arguments could be logically reversed and, for example, instead favor an evolving escape fraction on the basis of the present estimates of the star formation history.
More work is certainly needed in order to break such degeneracies, but a robust conclusion is that the simplest picture contains inconsistencies, the resolution of which is poised to teach us about early star formation, AGN activity, and perhaps more.

Better constraints on the thermal history of the IGM and on the abundance of the Lyman-limit systems which determine the mean free path of ionizing photons will be key in reducing the systematic uncertainties in the arguments presented here.
In addition, improved determinations of the spectral indices of quasars and star-forming galaxies, especially shortward of the Lyman limit, will be useful in refining our calculations.
In the same vein, an updated calculation of the full spectrum of the ionizing background, like that of \cite{1996ApJ...461...20H}, is clearly warranted in the light of our improved empirical knowledge of the physical state of the IGM and of the luminosity functions of both quasars and galaxies (Faucher-Gigu\`ere et al., in prep.)\nocite{ionspectrum}.

\acknowledgements
We thank Rychard Bouwens, Francesco Haardt, Andrew Hopkins, Philip Hopkins, Matthew McQuinn, Jason X. Prochaska, and Hy Trac for useful discussions.
We also thank the referee for a useful report which improved the presentation of this paper.
CAFG is supported by a NSERC Postgraduate Fellowship and a fellowship supplement from the Canadian Space Agency.
This work was supported in part by NSF grants ACI
96-19019, AST 00-71019, AST 02-06299, AST 03-07690, and AST 05-06556, and NASA ATP grants NAG5-12140, NAG5-13292, NAG5-13381, and NNG-05GJ40G.
Further support was provided by the David and Lucile Packard, the Alfred P. Sloan, and the John D. and Catherine T. MacArthur Foundations.

\bibliography{references}

\end{document}